\PassOptionsToPackage{usenames,dvipsnames,svgnames,table}{xcolor}
\documentclass[twocolumn, letter, natbib,twocolappendix,times,trackchanges]{aastex63_shp}
\pdfoutput=1
\usepackage{graphicx}
 \usepackage{xspace}
\usepackage{amsmath}
\usepackage{commath}
\usepackage{mathtools}
\usepackage{stix}
   
\usepackage{multirow,bigdelim}

\usepackage{enumitem}

\renewcommand{\baselinestretch}{0.935}

\makeatletter
\newcommand\makebig[2]{%
  \@xp\newcommand\@xp*\csname#1\endcsname{\bBigg@{#2}}%
  \@xp\newcommand\@xp*\csname#1l\endcsname{\@xp\mathopen\csname#1\endcsname}%
  \@xp\newcommand\@xp*\csname#1r\endcsname{\@xp\mathclose\csname#1\endcsname}%
}
\makeatother

\makebig{biggg} {3.0}
\makebig{Biggg} {3.5}
\makebig{bigggg}{4.0}
\makebig{Bigggg}{4.5}

\newcommand{\unit}[1]{\mathrm{\;#1}}
\newcommand{\Msun}{\ensuremath{M_{\odot}}\xspace}
\renewcommand{\AA}{\Angstrom}

\newcommand{\JWST}{\textit{JWST}}

\newcommand{\Halpha}{\ensuremath{\textrm{H}\alpha}\xspace}

\newcommand{\sigmaint}{\ensuremath{\sigma_0}\xspace}

\newcommand{\Mdyn}{\ensuremath{M_{\mathrm{dyn}}}\xspace}
\newcommand{\Mstar}{\ensuremath{M_*}\xspace}
\newcommand{\Mgas}{\ensuremath{M_{\mathrm{gas}}}\xspace}
\newcommand{\Mbar}{\ensuremath{M_{\mathrm{bar}}}\xspace}
\newcommand{\Mbulge}{\ensuremath{M_{\mathrm{bulge}}}\xspace}
\newcommand{\Mvir}{\ensuremath{M_{\mathrm{vir}}}\xspace}

\newcommand{\lMstar}{\ensuremath{\log_{10}(\Mstar/\Msun)}\xspace}
\newcommand{\lMgas}{\ensuremath{\log_{10}(\Mgas/\Msun)}\xspace}
\newcommand{\lMbar}{\ensuremath{\log_{10}(\Mbar/\Msun)}\xspace}
\newcommand{\lMvir}{\ensuremath{\log_{10}(\Mvir/\Msun)}\xspace}

\newcommand{\Sigbar}{\ensuremath{\Sigma_{\mathrm{bar}}}\xspace}
\newcommand{\lSigbar}{\ensuremath{\log_{10}(\Sigma_{\mathrm{bar}}/M_{\odot} \ \mathrm{kpc}^{-2})}\xspace}

\newcommand{\fgas}{\ensuremath{f_{\mathrm{gas}}}\xspace}

\newcommand{\re}{\ensuremath{R_e}\xspace}  
\newcommand{\rd}{\ensuremath{R_d}\xspace}  
\newcommand{\fDM}{\ensuremath{f_{\mathrm{DM}}(\re)}\xspace}
\newcommand{\fDMg}{\ensuremath{f_{\mathrm{DM}}}\xspace}

\newcommand{\rB}{\ensuremath{r_{\mathrm{B}}}\xspace}
\newcommand{\alphainner}{\ensuremath{\alpha_{\mathrm{inner}}}\xspace}

\newcommand{\vrot}{\ensuremath{v_{\mathrm{rot}}}\xspace}

\newcommand{\vcirc}{\ensuremath{v_{\mathrm{circ}}}\xspace}
\newcommand{\vcirctot}{\ensuremath{v_{\mathrm{circ,tot}}}\xspace}

\newcommand{\qint}{\ensuremath{q_{0,\mathrm{disk}}}\xspace}

\newcommand{\redisk}{\ensuremath{R_{e,\mathrm{disk}}}\xspace}  
\newcommand{\nSdisk}{\ensuremath{n_{S,\mathrm{disk}}}\xspace}  
\newcommand{\rebulge}{\ensuremath{R_{e,\mathrm{bulge}}}\xspace}  
\newcommand{\nSbulge}{\ensuremath{n_{S,\mathrm{bulge}}}\xspace}  

\newcommand{\bt}{\ensuremath{B/T}\xspace}  

\newcommand{\chalo}{\ensuremath{c_{\mathrm{halo}}}\xspace}

\newcommand{\alphaEinasto}{\ensuremath{\alpha_{\mathrm{Ein}}}\xspace}

\newcommand{\PA}{\ensuremath{\mathrm{PA}}\xspace}

\newcommand{\xgal}{\ensuremath{x_{\mathrm{gal}}}\xspace}
\newcommand{\ygal}{\ensuremath{y_{\mathrm{gal}}}\xspace}
\newcommand{\zgal}{\ensuremath{z_{\mathrm{gal}}}\xspace}
\newcommand{\ygalhat}{\ensuremath{\hat{y}_{\mathrm{gal}}}\xspace}

\newcommand{\Rgal}{\ensuremath{R_{\mathrm{gal}}}\xspace}

\newcommand{\Rout}{\ensuremath{R_{\mathrm{out}}}\xspace}

\newcommand{\xsky}{\ensuremath{x_{\mathrm{sky}}}\xspace}
\newcommand{\ysky}{\ensuremath{y_{\mathrm{sky}}}\xspace}
\newcommand{\zsky}{\ensuremath{z_{\mathrm{sky}}}\xspace}

\newcommand{\xoutf}{\ensuremath{x_{\mathrm{out}}}\xspace}
\newcommand{\youtf}{\ensuremath{y_{\mathrm{out}}}\xspace}
\newcommand{\zoutf}{\ensuremath{z_{\mathrm{out}}}\xspace}

\newcommand{\xcenter}{\ensuremath{x_{0}}\xspace}
\newcommand{\ycenter}{\ensuremath{y_{0}}\xspace}
\newcommand{\Vsys}{\ensuremath{V_{\mathrm{sys}}}\xspace}

\newcommand{\Vlos}{\ensuremath{V_{\mathrm{los}}}\xspace}

\newcommand{\Dysmalpy}{\texttt{DysmalPy}\xspace}

\defcitealias{Genzel20}{Paper~I}

\usepackage{etoolbox}

\makeatletter

\patchcmd{\NAT@citex}
  {\@citea\NAT@hyper@{%
     \NAT@nmfmt{\NAT@nm}%
     \hyper@natlinkbreak{\NAT@aysep\NAT@spacechar}{\@citeb\@extra@b@citeb}%
     \NAT@date}}
  {\@citea\NAT@nmfmt{\NAT@nm}%
   \NAT@aysep\NAT@spacechar\NAT@hyper@{\NAT@date}}{}{}

\patchcmd{\NAT@citex}
  {\@citea\NAT@hyper@{%
     \NAT@nmfmt{\NAT@nm}%
     \hyper@natlinkbreak{\NAT@spacechar\NAT@@open\if*#1*\else#1\NAT@spacechar\fi}%
       {\@citeb\@extra@b@citeb}%
     \NAT@date}}
  {\@citea\NAT@nmfmt{\NAT@nm}%
   \NAT@spacechar\NAT@@open\if*#1*\else#1\NAT@spacechar\fi\NAT@hyper@{\NAT@date}}
  {}{}

\makeatother
\journalinfo{Accepted for publication in the Astrophysical Journal}

\newcommand*{\MPE}{Max-Planck-Institut f\"{u}r extraterrestrische Physik (MPE), Giessenbachstr. 1, D-85748 Garching, Germany}
\newcommand*{\MPA}{Max-Planck-Institut f\"{u}r Astrophysik (MPA), Karl Schwarzschildstr. 1, D-85748 Garching, Germany}
\newcommand*{\UCB}{Departments of Physics and Astronomy, University of California, Berkeley, CA 94720, USA}
\newcommand*{\USM}{Universit\"{a}ts-Sternwarte Ludwig-Maximilians-Universit\"{a}t (USM), Scheinerstr. 1, D-81679 M\"{u}nchen, Germany}
\newcommand*{\TAU}{School of Physics and Astronomy, Tel Aviv University, Tel Aviv 69978, Israel}
\newcommand*{\CCA}{Center for Computational Astrophysics, Flatiron Institute, 162 5th Avenue, New York, NY 10010, USA}
\newcommand*{\UC}{Astronomy Department, Universidad de Concepci\'{o}n, Av. Esteban Iturra s/n Barrio Universitario, Casilla 160, Concepci\'{o}n, Chile}
\newcommand*{\IRAM}{Institute for Radio Astronomy in the Millimeter Range (IRAM), Rue de la Piscine, Grenoble, France}
\newcommand*{\INAF}{INAF - Osservatorio Astronomico di Padova, Vicolo dell'Osservatorio 5, I-35122 Padova, Italy}
\newcommand*{\ANU}{Research School of Astronomy and Astrophysics, Australian National University, Canberra, ACT 2611, Australia}
\newcommand*{\ASTROTD}{ARC Centre of Excellence for All Sky Astrophysics in 3 Dimensions (ASTRO 3D), Australia}
\newcommand*{\UB}{Department of Physics, University of Bath, Claverton Down, Bath BA2 7AY, UK}
\newcommand*{\Swinburne}{Centre for Astrophysics and Supercomputing, Swinburne University of Technology, John St, Hawthorn VIC 3122, Australia}

\begin{document}

\title{Rotation Curves in $\boldsymbol{z\sim1{-}2}$ Star-Forming Disks:\\
Comparison of Dark Matter Fractions and Disk Properties for Different Fitting Methods}
\shorttitle{Rotation Curves in $z\sim1{-}2$ Star-Forming Disks: Comparison of Fitting Methods}

\author{S. H. Price}\altaffiliation{Email: sedona@mpe.mpg.de}\affiliation{\MPE}
\author{T. T. Shimizu}\affiliation{\MPE}
\author{R. Genzel}\affiliation{\MPE}\affiliation{\UCB}
\author{H. \"{U}bler}\affiliation{\MPE}
\author{N. M. {F\"{o}rster Schreiber}}\affiliation{\MPE}
\author{L. J. Tacconi}\affiliation{\MPE}
\author{R. I. Davies}\affiliation{\MPE}
\author{R. T. Coogan}\affiliation{\MPE}
\author{D. Lutz}\affiliation{\MPE}
\author{S. Wuyts}\affiliation{\UB}
\author{E. Wisnioski}\affiliation{\ANU}\affiliation{\ASTROTD}
\author{A. Nestor}\affiliation{\TAU}
\author{A. Sternberg}\affiliation{\MPE}\affiliation{\TAU}\affiliation{\CCA}
\author{A. Burkert}\affiliation{\MPE}\affiliation{\USM}
\author{R. Bender}\affiliation{\MPE}\affiliation{\USM}
\author{A. Contursi}\affiliation{\MPE}\affiliation{\IRAM}
\author{R. L. Davies}\affiliation{\Swinburne}
\author{R. Herrera-Camus}\affiliation{\UC}
\author{M.-J. Lee}\affiliation{\MPE}
\author{T. Naab}\affiliation{\MPA}
\author{R. Neri}\affiliation{\IRAM}
\author{A. Renzini}\affiliation{\INAF}
\author{R. Saglia}\affiliation{\MPE}\affiliation{\USM}
\author{A. Schruba}\affiliation{\MPE}
\author{K. Schuster}\affiliation{\IRAM}

\shortauthors{Price et al.}

% ++++++++++++++++++++++++++++

\begin{abstract} 
We present a follow-up analysis examining the dynamics and structures of 41 massive, large star-forming galaxies 
at $z\sim0.67{-}2.45$ using both ionized and molecular gas kinematics. We fit the galaxy dynamics with models 
consisting of a bulge, a thick, turbulent disk, and a NFW dark matter halo, using code that fully forward models 
the kinematics, including all observational and instrumental effects. 
We explore the parameter space using Markov Chain Monte Carlo (MCMC) sampling, including priors based on 
stellar and gas masses and disk sizes. 
We fit the full sample using extracted 1D kinematic profiles. For a subset of 14 well-resolved galaxies, we also fit the 2D kinematics. 
The MCMC approach robustly confirms the results from least-squares fitting presented 
in \citetalias{Genzel20} \citep{Genzel20}: the sample galaxies tend to be baryon-rich 
on galactic scales (within one effective radius). 
The 1D and 2D MCMC results are also in good agreement for the subset, demonstrating that 
much of the galaxy dynamical information is captured along the major axis. 
The 2D kinematics are more affected by the presence of non-circular motions, which we illustrate 
by constructing a toy model with constant inflow for one galaxy that exhibits residual signatures consistent with radial motions. 
This analysis, together with results from \citetalias{Genzel20} and other studies, strengthens the finding 
that massive, star-forming galaxies at $z\sim1{-}2$ are baryon-dominated on galactic scales, 
with lower dark matter fractions towards higher baryonic surface densities. 
Finally, we present details of the kinematic fitting code used in this analysis.
\end{abstract}

%%%%%%%%%%

% ++++++++++++++++++++++++++++++++++++++++++++++++++++++++++++++++++++++++++++++++++++
% ++++++++++++++++++++++++++++++++++++++++++++++++++++++++++++++++++++++++++++++++++++

\section{Introduction}

Galaxy kinematics are a key probe of galaxy structure and mass distribution, as they provide 
a direct trace of the mass distribution that is not directly affected 
by dust attenuation or uncertainties in estimates of stellar or gas masses
(\citealt{vanderKruit78}, \citealt{vanderKruit11}, \citealt{Courteau14}).
Kinematic measurements can therefore be used to probe the amount of dark matter on galactic scales. 
Numerous studies over several decades have used kinematics to constrain 
the detailed mass distributions of nearby galaxies and their dark matter halos
(including \citealt{Rubin70}, \citealt{Freeman70}, \citealt{Casertano83}, \citealt{Carignan85}, \citealt{vanAlbada85}, and many others). 
In the local Universe, galaxy dynamics can be probed with stellar spectroscopy
and a wide range of gas tracers, including ionized, neutral, and molecular gas (\Halpha, \textsc{Hi}, and CO). 
For star-forming galaxies (SFGs), the gas tracers can often be used past the galaxies' optical extent 
(e.g. \citealt{Courteau14}, and references therein). 
Put together, kinematic measurements using all available tracers 
provide detailed constraints on local galaxies' mass distributions and dynamics, 
which tell us not only about the galaxies' present state, but also about how they formed.

Observations of galaxy kinematics at multiple epochs over cosmic time provide a powerful probe of the evolution of galactic structure, 
and of the relative amount and distribution of baryons versus dark matter. 
With the advent of sensitive near-infrared (near-IR) integral field unit (IFU) and slit spectrographs on 8-10m telescopes, 
ionized gas kinematics from strong rest-frame optical emission lines, in particular \Halpha, 
have become routinely accessible at $z\sim1{-}3$.  
Dynamical masses (\Mdyn) have been obtained for large samples of star-forming galaxies (SFGs) based on 
velocity and dispersion profiles in the central brighter regions, typically within the effective radius \re, 
near the peak of the rotation curve for an exponential disk ($R_{\mathrm{peak}}\sim1.3\re$).  
Several studies comparing \Mdyn estimates with stellar masses derived from multi-band photometry and 
gas masses obtained from observations of cold molecular gas/dust or inferred from scaling relations, 
have indicated that in these inner regions, 
$z\sim2$ SFGs have comparable or higher baryonic mass fractions than local disks of similar masses 
(e.g., \citealt{ForsterSchreiber06,ForsterSchreiber09}, \citealt{Erb06c}, \citealt{Price16,Price20}, \citealt{Wuyts16}).

Ideally, measurements should probe kinematics further out, well beyond $\sim1{-}1.5\,\re$ for two main reasons.  
Firstly, one can then use the \textit{shape} of the rotation curve (RC) to constrain the mass distribution, 
alleviating the large uncertainties associated with light-to-mass conversions in computing 
stellar (\Mstar) and gas masses (\Mgas). 
Secondly, the decomposition into baryonic and dark matter mass components is more robust 
when the range of radii probe from inner regions where baryons dominate into regions 
where the relative contribution of dark matter becomes more important.  
This requires very sensitive observations as the line emission from the galaxies gets very faint 
(e.g., exponential decrease of surface brightness with increasing radius for a disk with S\'ersic index $n_S=1$). 
Recent work capitalized on subsets with very deep data of individual galaxies, typically larger and/or with 
shallower light profile ($n_S\lesssim1$) than average, i.e. with measurable emission extending further into 
the (inner) dark matter halo (e.g., \citealt{Genzel17,Genzel20}, \citealt{Drew18}, \citealt{Ubler18}, \citealt{Molina19}). 
Other studies employed stacking techniques to derive the average RC of larger numbers of galaxies (\citealt{Lang17}, \citealt{Tiley19}). 
In yet another approach, \citet{vanDokkum15} constructed a composite RC, inferring rotation velocities 
from integrated line widths of a dozen compact and high-mass SFGs.  
Modeling of these individual or stacked RCs showed most robustly that $z\sim1{-}3$ SFGs tend to be 
strongly baryon-dominated on galactic scales (\citealt{Genzel17}, \citealt{Lang17}, \citealt{Ubler18}), 
although these findings were challenged by \citet{Tiley19}, who favored a different RC normalization scheme 
than \citet{Lang17} in their stacking approach.  
Mixed results in the literature likely reflect a combination of differences in methodologies and modeling, 
together with genuine trends among the galaxy population (e.g., see discussion by \citealt{ForsterSchreiber20}).

Modeling high-quality, individual extended \Halpha or CO RCs of 41 $z\sim0.67-2.45$ SFGs, 
\citet{Genzel20} recently confirmed the baryon dominance in a majority of these galaxies 
and also revealed important underlying trends, 
with \fDM anticorrelating most strongly with central mass surface density, 
bulge mass \Mbulge, and angular momentum parameter.  
These results echo trends highlighted from inner disk kinematics by \citet{Wuyts16} and 
seen in recent high-resolution numerical simulations of galaxy evolution (\citealt{Lovell18}, \citealt{Teklu18}, \citealt{Ubler21}). 
As shown by \citet{Genzel20}, low dark matter fractions on $\sim1\re$ scales can be explained by shallower inner 
dark matter mass distributions than ``NFW'' profiles \citep{Navarro96}, 
possibly due to heating by dynamical friction of satellites and AGN feedback.   
The results are also consistent with efficient radial transport in gas-rich environments at high redshift, 
leading to the rapid buildup of massive bulges and central black holes.

In this paper, we follow-up on \citet{Genzel20} (from here on, \citetalias{Genzel20}) by assessing the impact of fitting approach and of 
kinematic modeling in 1D versus 2D.  We model the same RCs of all 41 galaxies via a Markov Chain Monte Carlo (MCMC) technique 
to compare to the least-squares minimization adopted in \citetalias{Genzel20}.  
For a subset of 14 SFGs with deep adaptive optics-assisted IFU data, 
we further compare the results from modeling the 1D major-axis kinematic profiles to modeling the full 2D kinematic maps. 
Our analysis demonstrates the robustness of the findings of high baryonic fractions on galactic scales, 
and shows that the information about the overall mass distribution is well captured in the 1D major-axis profiles for these disk galaxies. 
We further highlight the potential of second-order deviations from axisymmetric rotation in probing the processes, such as radial inflows, 
that may efficiently concentrate baryons within the inner galactic regions at early $z\sim1{-}3$ epochs.  
We also present the details of the updated version of the galaxy kinematics modeling code \Dysmalpy.

Throughout, we assume a $\Lambda$CDM cosmology with $\Omega_m = 0.3$, $\Omega_{\Lambda} = 0.7$, 
and $H_0 = 70\unit{km\ s^{-1}\ Mpc^{-1}}$, and a \citet{Chabrier03} 
initial mass function.

%%%%%%%%%%%%%%%%%%%%%%%%%%%%%%%%%
%%%%%%%%%%%%%%%%%%%%%%%%%%%%%%%%%
\section{Sample and Data}
\label{sec:data}

\begin{deluxetable*}{lcc cccc c ccc}
\setlength{\tabcolsep}{0.04in} 
\renewcommand{\arraystretch}{1.05} 
\tabletypesize{\scriptsize}\tablenum{1}
\tablecaption{General Galaxy Parameters\tablenotemark{\tiny{a}}}
\label{tab:general}
\tablehead{
\colhead{ID} & \colhead{$z$} & \colhead{$\log_{10}(\Mstar/\Msun)_{\mathrm{SED}}$} & 
\colhead{SFR} &  \colhead{$\log_{10}(\Mgas/\Msun)$\tablenotemark{\tiny{b}}}  & 
\colhead{$\bt$} & 
\colhead{$\redisk^{0}$\tablenotemark{\tiny{c}}} & 
\colhead{$\nSdisk$} & \colhead{$\qint$} & \colhead{$i$} &  \colhead{$\chalo$}\\[-4pt]
\colhead{} & \colhead{} & \colhead{[dex]} & 
\colhead{[$\Msun\ \mathrm{yr}^{-1}$]} &  \colhead{[dex]}& 
\colhead{---} & 
\colhead{[kpc]} &  
\colhead{---} & \colhead{---} & \colhead{[deg]} & \colhead{---} 
} \startdata
EGS3\_10098 & 0.658 & 11.11 & 52.0 & 10.50 &  0.60 &  3.0  &  1.0 &  0.14 & 31.0 & 7.3 \\
U3\_21388 & 0.669 & 10.76 & 4.8 & 10.01 &  0.05 &  7.0  &  1.0 &  0.14 & 82.0 & 7.2 \\
EGS4\_21351 & 0.732 & 10.94 & 79.5 & 10.60 &  0.46 &  3.3  &  1.0 &  0.15 & 47.0 & 6.8 \\
EGS4\_11261 & 0.748 & 11.25 & 85.0 & 10.76 &  0.50 &  4.0  &  1.0 &  0.10 & 59.7 & 6.6 \\
GS4\_13143 & 0.760 & 9.82 & 12.0 & 9.82 &  0.70 &  5.6  &  1.3 &  0.15 & 74.0 & 6.5 \\
U3\_05138 & 0.809 & 10.20 & 6.0 & 9.88 &  0.50 &  7.5  &  1.0 &  0.12 & 55.0 & 6.5 \\
GS4\_03228 & 0.824 & 9.49 & 10.1 & 9.69 &  0.80 &  7.0  &  1.0 &  0.15 & 78.0 & 6.5 \\
GS4\_32976 & 0.831 & 10.37 & 22.0 & 10.19 &  0.90 &  6.8  &  1.0 &  0.17 & 68.0 & 6.5 \\
COS4\_01351 & 0.854 & 10.73 & 57.0 & 10.58 &  0.24 &  8.0  &  0.9 &  0.15 & 68.0 & 6.5 \\
COS3\_22796 & 0.914 & 10.32 & 11.1 & 10.10 &  0.15 &  9.0  &  1.0 &  0.17 & 58.0 & 6.1\\[2pt]
\arrayrulecolor{LightGrey}\hline
\arrayrulecolor{black}
U3\_15226 & 0.922 & 11.11 & 31.0 & 10.54 &  0.55 &  5.5  &  1.0 &  0.17 & 50.0 & 6.1 \\
GS4\_05881 & 0.990 & 9.78 & 19.0 & 9.92 &  0.85 &  5.6  &  1.3 &  0.17 & 60.0 & 6.0 \\
COS3\_16954 & 1.031 & 10.74 & 100.0 & 10.72 &  0.50 &  8.1  &  1.0 &  0.17 & 49.5 & 6.1 \\
COS3\_04796 & 1.032 & 10.80 & 51.0 & 10.63 &  0.18 &  9.7  &  1.1 &  0.10 & 50.0 & 6.1 \\
EGS\_13035123 & 1.120 & 11.18 & 126.0 & 10.95 &  0.20 &  10.2  &  1.0 &  0.17 & 24.0 & 6.0 \\
EGS\_13004291 & 1.197 & 10.97 & 630.0 & 11.54 &  0.61 &  3.0  &  1.3 &  0.17 & 27.0 & 6.0 \\
EGS\_13003805 & 1.232 & 11.23 & 200.0 & 11.32 &  0.29 &  5.6  &  1.2 &  0.17 & 37.0 & 5.9 \\
EGS4\_38153 & 1.362 & 10.44 & 78.0 & 10.55 &  0.16 &  5.9  &  1.0 &  0.20 & 75.0 & 5.0 \\
EGS4\_24985 & 1.400 & 10.90 & 99.0 & 10.70 &  0.40 &  4.6  &  1.0 &  0.20 & 40.0 & 5.0 \\
zC\_403741 & 1.446 & 10.65 & 60.0 & 10.45 &  0.68 &  2.6  &  1.0 &  0.20 & 28.0 & 5.0\\[2pt]
\arrayrulecolor{LightGrey}\hline
\arrayrulecolor{black}
D3a\_6397 & 1.500 & 11.08 & 214.0 & 11.00 &  0.57 &  6.3  &  1.0 &  0.24 & 30.0 & 5.0 \\
EGS\_13011166 & 1.530 & 11.08 & 375.0 & 11.41 &  0.55 &  6.3  &  1.0 &  0.20 & 60.0 & 5.0 \\
GS4\_43501 & 1.614 & 10.61 & 53.0 & 10.51 &  0.40 &  4.9  &  0.6 &  0.20 & 62.0 & 5.0 \\
GS4\_14152 & 1.615 & 11.30 & 167.0 & 11.07 &  0.23 &  6.8  &  1.0 &  0.20 & 55.0 & 5.0 \\
K20\_ID9 & 2.036 & 10.65 & 81.0 & 10.66 &  0.30 &  7.1  &  1.0 &  0.25 & 48.0 & 4.0 \\
zC\_405501 & 2.154 & 9.92 & 60.0 & 10.27 &  0.07 &  5.0  &  0.2 &  0.25 & 75.0 & 4.0 \\
SSA22\_MD41 & 2.172 & 9.86 & 130.0 & 10.46 &  0.05 &  7.1  &  0.4 &  0.25 & 72.0 & 4.0 \\
BX389 & 2.180 & 10.60 & 100.0 & 10.68 &  0.30 &  7.4  &  0.2 &  0.25 & 76.0 & 4.0 \\
zC\_407302 & 2.182 & 10.39 & 340.0 & 10.60 &  0.50 &  4.0  &  1.0 &  0.12 & 60.0 & 4.0 \\
GS3\_24273 & 2.187 & 11.00 & 267.0 & 11.03 &  0.80 &  7.0  &  1.0 &  0.25 & 60.0 & 4.0\\[2pt]
\arrayrulecolor{LightGrey}\hline
\arrayrulecolor{black}
zC\_406690 & 2.196 & 10.62 & 300.0 & 10.91 &  0.90 &  4.5  &  0.2 &  0.25 & 25.0 & 4.0 \\
BX610 & 2.210 & 11.00 & 60.0 & 11.36 &  0.42 &  4.9  &  1.0 &  0.25 & 39.0 & 4.0 \\
K20\_ID7 & 2.225 & 10.28 & 101.0 & 10.59 &  0.03 &  8.2  &  0.2 &  0.25 & 64.0 & 4.0 \\
K20\_ID6 & 2.236 & 10.43 & 99.0 & 10.57 &  0.30 &  5.0  &  0.5 &  0.25 & 31.0 & 4.0 \\
zC\_400569 & 2.242 & 11.08 & 240.0 & 11.30 &  0.70 &  4.0  &  1.0 &  0.22 & 45.0 & 4.0 \\
BX482 & 2.258 & 10.26 & 80.0 & 10.91 &  0.02 &  5.8  &  0.2 &  0.25 & 60.0 & 4.0 \\
COS4\_02672 & 2.308 & 10.57 & 72.0 & 10.61 &  0.10 &  7.4  &  0.5 &  0.25 & 62.0 & 4.0 \\
D3a\_15504 & 2.383 & 11.04 & 146.0 & 10.92 &  0.30 &  6.1  &  1.0 &  0.25 & 40.0 & 4.0 \\
D3a\_6004 & 2.387 & 11.50 & 355.0 & 11.27 &  0.44 &  5.3  &  0.4 &  0.25 & 20.0 & 4.0 \\
GS4\_37124 & 2.431 & 10.59 & 194.0 & 10.70 &  0.70 &  3.2  &  1.0 &  0.25 & 67.0 & 4.0 \\
GS4\_42930 & 2.451 & 10.33 & 70.0 & 10.37 &  0.50 &  2.8  &  1.2 &  0.25 & 59.0 & 4.0\\[2pt]
\enddata
\tablenotetext{a}{Parameters include redshift ($z$), stellar mass (\Mstar), SFR, gas mass (\Mgas), bulge-to-total ratio (\bt), 
the effective radius (\redisk), S\'ersic index (\nSdisk), and intrinsic axis ratio (\qint) of the disk, 
inclination ($i$), and halo concentration (\chalo).}\vspace{-6pt}
\tablenotetext{b}{From direct measurements, or gas-mass scaling relations.}\vspace{-6pt}
\tablenotetext{c}{Best-fit \redisk from \citetalias{Genzel20}.}
\vspace{-10pt} 
\end{deluxetable*}

The RC41 sample consists of 41 individual star-forming and kinematically-classified disk galaxies at redshifts $z=0.65{-}2.45$.
As shown in Figure~1 of \citetalias{Genzel20}, these galaxies range 
in stellar mass and size from $\lMstar \sim 9.8{-}11.4$ and $\re \sim 2.5{-}10\unit{kpc}$ 
(the projected major-axis half-light radius from stellar rest-frame $5000\mathrm{\AA}$ light). 
The galaxies lie near the SFR versus \Mstar ``main sequence'' of SFGs (MS; e.g.,  \citealt{Speagle14}, \citealt{Whitaker14a}), 
and tend to be larger than average at their stellar mass and redshift (based on the mass-size relationship of \citealt{vanderWel14a}).  
The size bias results mainly from the selection of galaxies with sufficiently well-resolved data and 
the most extended kinematic profiles to enable analysis of the outer disk RCs (i.e. beyond $\sim1{-}1.5\,\re$).

The kinematic data are drawn from observations of ionized gas (traced with \Halpha) and cold molecular gas (traced with CO). 
The data come from near infrared observations with SINFONI (IFU, in both seeing-limited and adaptive optics-assisted modes) 
and KMOS (IFU, seeing-limited) at ESO/VLT, and with LBT-LUCI (slit, seeing-limited), 
together with millimeter interferometry from IRAM/NOEMA. 
The observations for each galaxy probe out to $\sim\!1.5{-}4$ times the effective radius, 
which is crucial for detailed kinematic modeling. 
The median on-source integration time is 16 hours, with individual data sets ranging from 4 to 56 hours.
Table~\ref{tab:general} lists the basic properties of the sample, including $z$, \Mstar, SFR, gas mass, and 
structural parameters, as well as the halo concentration parameter 
(adopted as a function of $z$ only, based on the average of relations by \citealt{Bullock01}, 
\citealt{Dutton14}, \citealt{Ludlow14}; see Tables~1, D1, D2 and Sec.~2 and Appendix~A of \citetalias{Genzel20} for full details).

Most of the information about the underlying mass distribution of disks is expected to be encoded in 
the velocity and velocity dispersion profiles along the projected kinematic major axis 
(c.f., \citealt{Genzel06, Genzel17}; see also Section~\ref{sec:comp1D2D}).  
For the comparison of fitting methodology, we use the same 1D profiles as in \citetalias{Genzel20}, 
where details of their extraction can be found.  
In summary, except for the LBT/LUCI slit spectroscopic \Halpha data of two objects, 
the profiles were obtained from the \Halpha or CO data cubes using a pseudo-slit 
along the kinematic major axis. 
The pseudo-slit width is either constant and $\sim\!1{-}1.5$ times the point-spread function full-width at half-maximum (PSF FWHM), 
or increasing towards the outskirts for more face-on galaxies (with opening angle $\sim\!5{-}10$ degrees).  
This choice of slit width and geometry best samples the major-axis kinematics given the projected isovelocity contours of inclined disks, 
maximizing the S/N and radial extent while minimizing contamination of signal away from the major axis. 
For galaxies with IFU observations in more than one mode (seeing-limited or AO-assisted) or more than one instrument 
(SINFONI + KMOS), the data are combined into a composite cube before performing the 1D extraction.  
For EGS4-24985 and EGS-13011166, observed with LBT/LUCI and with NOEMA, 
the independently extracted 1D profiles are combined together (see also \citealt{Genzel13}, \citealt{Ubler18}).

For the subset of galaxies with AO observations, we also extract 2D velocity and dispersion maps. 
As most of these objects also have seeing-limited observations, 
we extract maps from the cubes combining the AO and non-AO \Halpha data 
in order to maximize the S/N and radial coverage. 
For the composite cubes, uncertainties are estimated using RMS flux variations in the spectral ranges free from emission lines. 
We then use LINEFIT to measure the 2D kinematic maps from the composite cubes 
(\citealt{ForsterSchreiber09,ForsterSchreiber18}, \citealt{Davies11}). 
The velocity and dispersion maps (corrected for instrumental spectral resolution) 
are obtained by fitting a Gaussian to the \Halpha line in the spectrum of each spaxel, 
after spectral and spatial median-filtering of the input data cube by 3 pixels in each dimension 
(similar to the resolution element, to mitigate noise peaks).  
Uncertainties are derived through a Monte Carlo approach, by perturbing the input spectrum 100 times 
according to the noise cube associated with each data cube and assuming a Gaussian noise distribution. 
Masks for the 2D maps are created using a combination of criteria. 
These include an integrated line flux S/N cut (generally $\gtrsim\!\!5\sigma$), 
together with (where necessary) an integrated line flux fraction cut ($\gtrsim \! 0.05{-}0.1 f_{\mathrm{max}}$). 
We also use fit uncertainty cuts for both the velocity and dispersion maps (typically $\gtrsim\!3{-}5\sigma$)  
as well as sigma clipping of the kinematics as needed (mostly $\gtrsim 3{-}5\sigma$). 
Segmentation maps are also incorporated to detangle the flux from e.g.\ neighboring bright clumps. 
We additionally flag a small number of outlier pixels not otherwise excluded, 
or explicitly include a few non-problematic lower flux pixels falling within otherwise unmasked regions. 
In order to have the highest possible spatial resolution, when examining features in 
residual maps we additionally consider only the AO data (see Section~\ref{sec:comp1D2D_noncirc}).

\vspace{10pt}

%%%%%%%%%%%%%%%%%%%%%%%%%%%%%%%%%
%%%%%%%%%%%%%%%%%%%%%%%%%%%%%%%%%
\section{Dynamical Modeling}  
\label{sec:dyn_model}

To model the kinematics of the galaxies, we employ the fully 3D code DYSMAL, 
which has been continually optimized for applications to high redshift studies 
(\citealt{Genzel06,Genzel11,Genzel14,Genzel17,Genzel20}, \citealt{Cresci09}, \citealt{Davies11}, 
\citealt{Wuyts16}, \citealt{Lang17}, \citealt{Ubler18}). 
The code is parametric, and follows a forward modeling approach, incorporating one or more mass and kinematic components 
and accounting for all observational effects (such as projection, beam smearing, etc.).  
The fits to 1D and 2D measurements use profiles and maps from the 3D model cube 
extracted in a similar way as for the observations. 
The latest DYSMAL upgrade in functionality and model ingredients is presented in Appendix~\ref{sec:appendixA}. 
In this section, we give the specific model choices used for the present analysis.

\vspace{6pt}

% +++++++++++++++++++++++++++++++++++++++++++++++++++++++
\subsection{Galaxy Model Components}
\label{sec:model_gal_comp}

As in \citetalias{Genzel20}, we model each galaxy as a thick, turbulent disk and a bulge 
embedded in a dark matter halo, adopting the same treatment and assumptions.  
The disk and bulge components are modeled as deprojected S\'ersic profiles, following \citet{Noordermeer08}. 
We assume the bulge is spherical, with index $\nSbulge=4$ and projected effective radius $\rebulge = 1\unit{kpc}$. 
The disk is taken to be an oblate, flattened spheroid with intrinsic axis ratio \qint, index \nSdisk, and radius \redisk. 
The masses of the disk and bulge components are calculated based on 
the total baryonic mass, \lMbar, and the bulge-to-total ratio, \bt. 
The circular velocity curve for the disk and bulge are then calculated using Eq.~10 of \citet{Noordermeer08}. 
Following the approach in \citetalias{Genzel20} for consistency, we assume that only the disk component 
contributes to the light of our tracer (\Halpha or CO), and that the disk light distribution follows a S\'ersic profile.

Our model also includes a dark matter halo with a NFW profile \citep{Navarro96}, 
to provide the most direct comparison to the least-squares analysis in \citetalias{Genzel20}.
The halo has a mass $\Mvir$ and fixed concentration $\chalo$, which is adopted based on the 
typical redshift evolution of dark matter halo concentrations (e.g., \citealt{Bullock01}, \citealt{Dutton14}, \citealt{Moster20}). 
For this analysis, we do not include any halo adiabatic contraction.

We assume the intrinsic velocity dispersion of our galaxies is constant and isotropic throughout the disk, with value \sigmaint. 
The corresponding pressure support --- important for the dynamics of thick disks --- is accounted for by applying 
an asymmetric drift correction to the model circular velocity \vcirc in computing the actual rotation velocity \vrot following 
the relation presented by \citet{Burkert10} for exponential disks (their Eq.~11; see also Eq.~\ref{eq:asymm_drift}).

\begin{deluxetable}{c l l}[!h]
\setlength{\tabcolsep}{0.06in}
\tabletypesize{\scriptsize}\tablenum{2}
\tablecaption{Priors for 1D MCMC Fitting}
\label{tab:1d_mcmc_fit_priors}
\tablehead{
\colhead{Parameter} & \colhead{Prior} & \colhead{Bounds}  
}  \startdata\\[-8pt]
\lMbar	& Gaus$\left(\log_{10}(M_{*,\mathrm{SED}}+\Mgas),\ 0.2\unit{dex}\right)$\tablenotemark{\tiny{a}} 	& $[9, 13]\unit{dex}$  \\
\fDM 	& Flat  & $[0,1]$  \\
\sigmaint 	& Flat  & $[5,300]\unit{km\ s^{-1}}$  \\
\redisk 	& Gaus$\left(R^0_{E,\mathrm{disk}}, 2\unit{kpc}\right)$\tablenotemark{\tiny{b}}  & $[1,15]\unit{kpc}$\\[4pt]
\enddata
\tablenotetext{a}{$\log_{10}(M_{*}/M_{\odot})_{\mathrm{SED}}$ and \lMgas are listed in Table~\ref{tab:general}.}
\vspace{-6pt}
\tablenotetext{b}{The adopted values of $R^0_{E,\mathrm{disk}}$ are given in Table~\ref{tab:general}.}
\vspace{-20pt}
\end{deluxetable}

\vspace{6pt}

% +++++++++++++++++++++++++++++++++++++++++++++++++++++++
\subsection{Fitting 1D Kinematic Profiles}
\label{sec:model_fitting_1D}

The best-fit modeling parameters in \citetalias{Genzel20} were derived from least-squares optimization, 
using boundaries on the free parameters that are based on stellar, gas, and morphological properties. 
Here we instead use a MCMC parameter space exploration to determine the best-fit parameter values, 
following the general procedure described in Appendix~\ref{sec:DYSMALfitting}. 
The key difference is that the best-fit values for \citetalias{Genzel20} 
are determined using $\chi^2$ minimization (i.e., maximizing the model likelihood given the data),\footnote{The \citetalias{Genzel20} 
\fDM uncertainties are derived from a comparable MCMC fit (i.e., same free parameters), 
as MCMC sampling can efficiently and robustly capture multidimensional uncertainties (see \citetalias{Genzel20}, Appendix A.4).} 
while for this analysis we use the MCMC-derived posterior distributions to determine best-fit values 
(i.e., combining the prior and likelihood information).

In this analysis, we simultaneously fit the 1D velocity and dispersion profiles for 4 free parameters: 
the log total baryonic mass, \lMbar; the disk effective radius, \redisk; 
the intrinsic velocity dispersion, \sigmaint; and the enclosed dark matter fraction $\fDM=v_{\mathrm{circ,DM}}^2(\re)/\vcirctot^2(\re)$ 
(where here $\re=\redisk$\footnote{See also the discussion in Appendix~\ref{sec:MCMCprior} for the choice of 
fitting parameter and effective priors on quantities inferred from fitted parameters, 
in particular the choice of fitting \fDM versus \lMvir.}). 
We assume Gaussian priors for \lMbar with a standard deviation of $0.2\unit{dex}$ that are centered on 
the baryonic mass derived using \lMstar from SED fitting and either a direct measurement of \lMgas{}
(from \citealt{Tacconi13,Tacconi18}, and \citealt{Freundlich19}) 
or an estimate using the scaling relations from \citet{Tacconi20}. 
We also bound the values within $\lMbar\in[9,13]\unit{dex}$. 
For \redisk, we adopt Gaussian priors of standard deviation $2\unit{kpc}$ that are 
centered on the fit value of \redisk from \citetalias{Genzel20},\footnote{While this prior choice is not independent from our data,
it does allow us to include more information than just morphological fitting to imaging.} 
and also bound $\redisk\in[1,15]\unit{kpc}$. 
Finally, we adopt flat bounded priors for the intrinsic dispersion ($\sigmaint\in[5,300]\unit{km\ s^{-1}}$) 
and dark matter fraction ($\fDM\in[0,1]$). 
For reference, the priors adopted for the 1D fitting are summarized in Table~\ref{tab:1d_mcmc_fit_priors}.

The remaining model parameters are fixed, as it is difficult to simultaneously fit for more parameters 
given the spatial resolution and S/N of the data. 
We adopt the values of \nSdisk,  halo concentration \chalo, and the disk flattening \qint used in \citetalias{Genzel20} 
(see Tables~D1~\&~D2). 
We also use the final values of inclination $i$ and \bt from \citetalias{Genzel20} 
(determined from a combination of kinematic and imaging information), 
because estimates of inclination or \bt based only on imaging 
may also suffer from the effects of attenuation or mass-to-light gradients. 
The model position angle \PA is set to the measured kinematic major axis \PA (as in \citetalias{Genzel20}; 
very similar to the rest-frame optical morphological \PA). 
As the 1D velocity and dispersion profiles have been centered spatially and 
corrected for any systemic velocity, we fix $x_0=y_0=\Vsys=0$.

\begin{deluxetable*}{lc c ccc c}[!hbt]
\setlength{\tabcolsep}{0.06in} 
\renewcommand{\arraystretch}{1.1}
\tabletypesize{\scriptsize}\tablenum{3}
\tablecaption{Best-fit Parameters, 1D Fitting\tablenotemark{\tiny{a}}}
\label{tab:results_NFW}
\tablehead{
\colhead{ID} & \colhead{$z$} & \colhead{\lMbar} & 
\colhead{\re} & \colhead{\sigmaint} & 
\colhead{\fDM} & \colhead{\lMvir\tablenotemark{\tiny{b}}}\\[-4pt]
\colhead{} & \colhead{} & \colhead{[dex]} & 
\colhead{[kpc]} & \colhead{[$\mathrm{km\ s^{-1}}$]} & 
\colhead{---} & \colhead{[dex]}\\[-6pt]
\colhead{} & \colhead{} & \colhead{\textit{[Free]}} & 
\colhead{\textit{[Free]}}
&  \colhead{\textit{[Free]}} & 
\colhead{\textit{[Free]}} & \colhead{\textit{[Inferred]}} 
 }  \startdata
EGS3\_10098 & 0.658 & $10.93_{-0.13}^{+0.11}$& $3.93_{-1.18}^{+1.95}$& $61.79_{-15.07}^{+9.99}$& $0.22_{-0.22}^{+0.13}$& $12.55_{-0.72}^{+1.21}$ \\
U3\_21388 & 0.669 & $10.80_{-0.19}^{+0.17}$& $8.68_{-1.42}^{+1.52}$& $51.04_{-6.88}^{+7.78}$& $0.69_{-0.09}^{+0.13}$& $12.72_{-0.16}^{+0.22}$ \\
EGS4\_21351 & 0.732 & $10.68_{-0.14}^{+0.07}$& $5.90_{-1.52}^{+1.71}$& $30.00_{-6.27}^{+7.01}$& $0.11_{-0.11}^{+0.13}$& $10.53_{-0.36}^{+1.26}$ \\
EGS4\_11261 & 0.748 & $11.11_{-0.07}^{+0.20}$& $5.49_{-1.33}^{+2.75}$& $41.34_{-11.22}^{+8.46}$& $0.24_{-0.14}^{+0.11}$& $12.52_{-0.70}^{+0.78}$ \\
GS4\_13143 & 0.760 & $10.25_{-0.10}^{+0.15}$& $5.27_{-1.31}^{+1.92}$& $21.70_{-8.06}^{+6.37}$& $0.48_{-0.13}^{+0.13}$& $11.75_{-0.26}^{+0.26}$ \\
U3\_05138 & 0.809 & $10.41_{-0.12}^{+0.12}$& $7.57_{-1.54}^{+2.06}$& $16.88_{-9.41}^{+3.79}$& $0.45_{-0.16}^{+0.21}$& $11.34_{-0.33}^{+0.49}$ \\
GS4\_03228 & 0.824 & $9.98_{-0.14}^{+0.19}$& $6.76_{-2.01}^{+1.76}$& $13.88_{-8.58}^{+3.83}$& $0.75_{-0.12}^{+0.11}$& $11.88_{-0.16}^{+0.20}$ \\
GS4\_32976 & 0.831 & $10.71_{-0.08}^{+0.14}$& $9.40_{-1.32}^{+2.44}$& $40.53_{-9.70}^{+6.53}$& $0.62_{-0.10}^{+0.08}$& $12.52_{-0.25}^{+0.26}$ \\
COS4\_01351 & 0.854 & $10.97_{-0.09}^{+0.18}$& $7.08_{-1.04}^{+2.63}$& $63.67_{-4.96}^{+5.09}$& $0.53_{-0.09}^{+0.11}$& $12.87_{-0.26}^{+0.23}$ \\
COS3\_22796 & 0.914 & $10.54_{-0.15}^{+0.18}$& $9.56_{-2.01}^{+1.50}$& $11.00_{-5.92}^{+6.74}$& $0.58_{-0.18}^{+0.20}$& $11.57_{-0.33}^{+0.38}$\\[2pt]
\arrayrulecolor{LightGrey}\hline
\arrayrulecolor{black}
U3\_15226 & 0.922 & $10.67_{-0.09}^{+0.15}$& $6.65_{-1.47}^{+2.52}$& $42.19_{-7.87}^{+5.26}$& $0.42_{-0.16}^{+0.10}$& $11.96_{-0.43}^{+0.33}$ \\
GS4\_05881 & 0.990 & $10.13_{-0.12}^{+0.13}$& $4.71_{-1.79}^{+2.31}$& $63.48_{-6.60}^{+5.11}$& $0.76_{-0.07}^{+0.12}$& $12.92_{-0.32}^{+0.18}$ \\
COS3\_16954 & 1.031 & $10.87_{-0.09}^{+0.13}$& $8.20_{-1.33}^{+2.21}$& $55.10_{-7.23}^{+5.29}$& $0.63_{-0.10}^{+0.12}$& $12.90_{-0.22}^{+0.21}$ \\
COS3\_04796 & 1.032 & $11.12_{-0.09}^{+0.14}$& $9.22_{-1.44}^{+2.18}$& $18.82_{-10.80}^{+2.48}$& $0.49_{-0.16}^{+0.12}$& $12.54_{-0.34}^{+0.27}$ \\
EGS\_13035123 & 1.120 & $11.09_{-0.09}^{+0.09}$& $10.17_{-1.73}^{+1.92}$& $19.33_{-1.89}^{+2.09}$& $0.28_{-0.16}^{+0.15}$& $11.45_{-0.42}^{+0.56}$ \\
EGS\_13004291 & 1.197 & $11.12_{-0.08}^{+0.05}$& $4.48_{-0.87}^{+2.15}$& $59.34_{-5.26}^{+8.83}$& $0.08_{-0.08}^{+0.16}$& $11.35_{-0.06}^{+1.66}$ \\
EGS\_13003805 & 1.232 & $11.43_{-0.11}^{+0.09}$& $6.70_{-1.76}^{+1.61}$& $40.45_{-10.15}^{+9.86}$& $0.18_{-0.17}^{+0.09}$& $12.19_{-0.55}^{+1.08}$ \\
EGS4\_38153 & 1.362 & $10.94_{-0.20}^{+0.17}$& $4.17_{-1.36}^{+1.91}$& $58.48_{-17.44}^{+8.70}$& $0.47_{-0.15}^{+0.20}$& $13.50_{-0.25}^{+0.38}$ \\
EGS4\_24985 & 1.400 & $11.14_{-0.13}^{+0.12}$& $5.94_{-1.40}^{+1.39}$& $43.03_{-27.70}^{+7.25}$& $0.35_{-0.16}^{+0.17}$& $12.83_{-0.44}^{+0.50}$ \\
zC\_403741 & 1.446 & $10.60_{-0.10}^{+0.04}$& $3.28_{-0.67}^{+2.27}$& $69.48_{-8.08}^{+6.27}$& $0.05_{-0.05}^{+0.18}$& $10.29_{-0.07}^{+1.86}$\\[2pt]
\arrayrulecolor{LightGrey}\hline
\arrayrulecolor{black}
D3a\_6397 & 1.500 & $11.13_{-0.07}^{+0.08}$& $6.83_{-1.68}^{+1.96}$& $82.52_{-11.02}^{+7.17}$& $0.38_{-0.17}^{+0.14}$& $12.70_{-0.46}^{+0.54}$ \\
EGS\_13011166 & 1.530 & $11.25_{-0.08}^{+0.11}$& $7.80_{-1.42}^{+2.03}$& $60.93_{-8.67}^{+7.61}$& $0.34_{-0.10}^{+0.12}$& $12.53_{-0.33}^{+0.44}$ \\
GS4\_43501 & 1.614 & $10.82_{-0.11}^{+0.16}$& $5.05_{-1.28}^{+2.15}$& $46.04_{-8.45}^{+5.16}$& $0.38_{-0.10}^{+0.12}$& $12.31_{-0.27}^{+0.37}$ \\
GS4\_14152 & 1.615 & $11.45_{-0.10}^{+0.13}$& $7.35_{-1.67}^{+1.69}$& $45.53_{-11.61}^{+8.12}$& $0.30_{-0.15}^{+0.13}$& $12.69_{-0.49}^{+0.56}$ \\
K20\_ID9 & 2.036 & $10.92_{-0.11}^{+0.16}$& $7.00_{-1.31}^{+2.19}$& $26.64_{-13.07}^{+5.03}$& $0.46_{-0.17}^{+0.17}$& $12.23_{-0.38}^{+0.36}$ \\
zC\_405501 & 2.154 & $10.46_{-0.16}^{+0.22}$& $6.03_{-1.07}^{+1.34}$& $65.68_{-5.99}^{+4.02}$& $0.52_{-0.21}^{+0.21}$& $11.41_{-0.41}^{+0.44}$ \\
SSA22\_MD41 & 2.172 & $10.61_{-0.18}^{+0.24}$& $7.98_{-1.27}^{+1.70}$& $72.20_{-6.85}^{+6.52}$& $0.71_{-0.14}^{+0.15}$& $12.05_{-0.24}^{+0.25}$ \\
BX389 & 2.180 & $11.05_{-0.11}^{+0.27}$& $6.81_{-0.87}^{+1.67}$& $80.33_{-8.27}^{+4.94}$& $0.59_{-0.15}^{+0.14}$& $13.06_{-0.34}^{+0.31}$ \\
zC\_407302 & 2.182 & $10.66_{-0.10}^{+0.11}$& $5.39_{-1.54}^{+1.82}$& $63.62_{-6.26}^{+5.29}$& $0.61_{-0.09}^{+0.10}$& $12.81_{-0.24}^{+0.24}$ \\
GS3\_24273 & 2.187 & $10.91_{-0.08}^{+0.08}$& $8.80_{-1.23}^{+1.90}$& $21.78_{-9.48}^{+7.65}$& $0.25_{-0.10}^{+0.15}$& $11.18_{-0.30}^{+0.55}$\\[2pt]
\arrayrulecolor{LightGrey}\hline
\arrayrulecolor{black}
zC\_406690 & 2.196 & $11.06_{-0.06}^{+0.04}$& $4.84_{-0.92}^{+2.26}$& $73.39_{-3.11}^{+7.14}$& $0.06_{-0.06}^{+0.12}$& $10.81_{-0.18}^{+1.46}$ \\
BX610 & 2.210 & $11.06_{-0.10}^{+0.11}$& $6.02_{-1.28}^{+2.02}$& $80.39_{-6.41}^{+5.68}$& $0.45_{-0.13}^{+0.16}$& $12.80_{-0.37}^{+0.45}$ \\
K20\_ID7 & 2.225 & $10.89_{-0.20}^{+0.24}$& $7.70_{-1.30}^{+1.40}$& $74.30_{-6.97}^{+4.44}$& $0.76_{-0.10}^{+0.16}$& $13.02_{-0.14}^{+0.22}$ \\
K20\_ID6 & 2.236 & $10.69_{-0.13}^{+0.10}$& $4.88_{-0.78}^{+2.61}$& $64.70_{-5.33}^{+7.45}$& $0.13_{-0.13}^{+0.20}$& $10.52_{-0.21}^{+1.23}$ \\
zC\_400569 & 2.242 & $10.98_{-0.04}^{+0.09}$& $6.18_{-1.51}^{+2.15}$& $71.82_{-8.40}^{+5.97}$& $0.29_{-0.18}^{+0.12}$& $11.92_{-0.73}^{+0.75}$ \\
BX482 & 2.258 & $11.05_{-0.15}^{+0.23}$& $6.58_{-0.79}^{+0.98}$& $71.63_{-4.03}^{+3.33}$& $0.61_{-0.16}^{+0.17}$& $12.93_{-0.27}^{+0.30}$ \\
COS4\_02672 & 2.308 & $10.85_{-0.10}^{+0.22}$& $7.08_{-1.24}^{+1.97}$& $64.67_{-4.51}^{+4.07}$& $0.43_{-0.21}^{+0.20}$& $11.61_{-0.42}^{+0.47}$ \\
D3a\_15504 & 2.383 & $11.13_{-0.08}^{+0.12}$& $6.40_{-1.30}^{+1.79}$& $69.84_{-4.75}^{+3.95}$& $0.24_{-0.10}^{+0.13}$& $11.68_{-0.41}^{+0.59}$ \\
D3a\_6004 & 2.387 & $11.52_{-0.11}^{+0.07}$& $6.05_{-1.08}^{+2.08}$& $68.68_{-6.36}^{+6.68}$& $0.12_{-0.12}^{+0.17}$& $11.93_{-0.26}^{+1.49}$ \\
GS4\_37124 & 2.431 & $10.78_{-0.11}^{+0.15}$& $5.17_{-1.23}^{+1.86}$& $68.48_{-13.84}^{+7.92}$& $0.43_{-0.15}^{+0.11}$& $12.33_{-0.54}^{+0.42}$ \\
GS4\_42930 & 2.451 & $10.43_{-0.07}^{+0.15}$& $5.17_{-1.31}^{+1.85}$& $56.91_{-4.35}^{+3.39}$& $0.43_{-0.18}^{+0.16}$& $11.35_{-0.61}^{+0.61}$\\[2pt]
\enddata
\tablenotetext{a}{MCMC MAP values (from the joint posterior distribution) for fits to 1D data, 
using NFW halos, no adiabatic contraction, and assuming asymmetric drift corrections. }
\vspace{-6pt}
\tablenotetext{b}{Calculated from the best-fit \fDM, \lMbar, and \re.}
\vspace{-20pt}
\end{deluxetable*}

% ++++++++++++++++++++++++++++
% ++++++++++++++++++++++++++++
\begin{figure*}[ht!]
\vglue -4pt
\centering
\hglue -4.5pt
\includegraphics[width=1.017\textwidth]{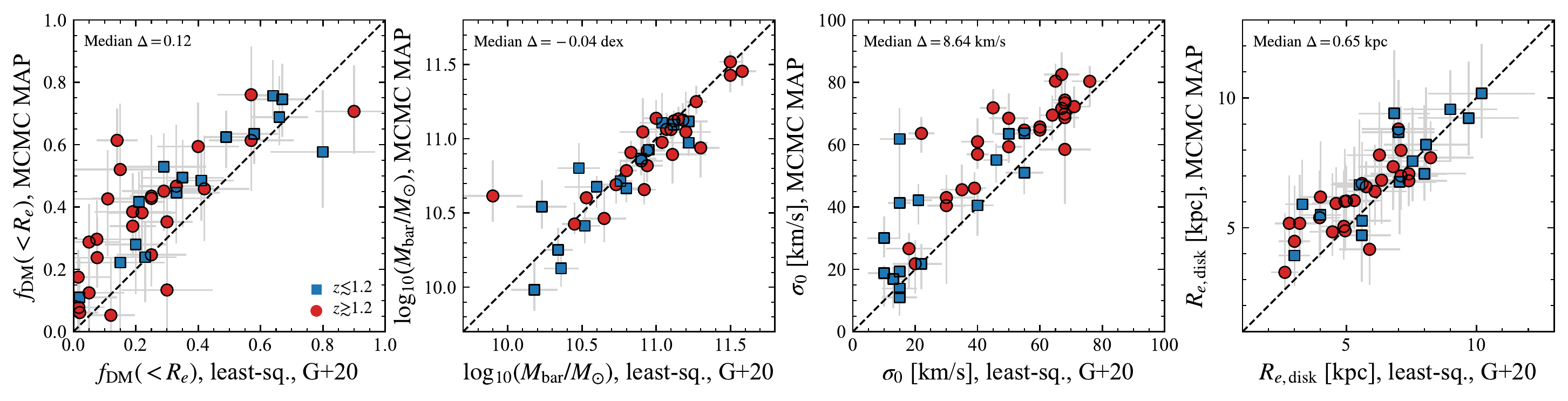}
\vglue -6pt
\caption{
Comparison of the best-fit values of the key dynamical and structural parameters for the RC41 sample 
from \citetalias{Genzel20} (least squares) and the 1D MCMC analysis (MAP) presented in this paper. 
From left to right, we compare \fDM, \lMbar, \sigmaint, and \redisk. 
Galaxies at $z<1.2$ and $z\geq1.2$ are marked with blue squares and red circles, respectively. 
The black dashed line shows the 1:1 relation. 
The dark matter fractions measured from the MCMC MAP analysis in this paper tend to be higher than 
those from \citetalias{Genzel20} by $\Delta\fDM\sim0.1{-}0.15$, 
but the values are consistent within the uncertainties for most galaxies. 
Paired with the \fDM offset, in this paper we tend to find slightly lower baryonic mass values. 
We tend to find higher intrinsic velocity dispersion values, as the priors adopted in this analysis are 
\textit{less restrictive} than the bounds in \citetalias{Genzel20}. 
The best-fit disk effective radii are fairly similar between the two analyses, though slightly larger in this paper's analysis. 
}
\label{fig:1D_chisq_vs_mcmc}
\end{figure*}
% ++++++++++++++++++++++++++++
% ++++++++++++++++++++++++++++

% ++++++++++++++++++++++++++++
\begin{figure*}[!htb]
\centering
\vglue -12pt
\includegraphics[width=0.75\textwidth]{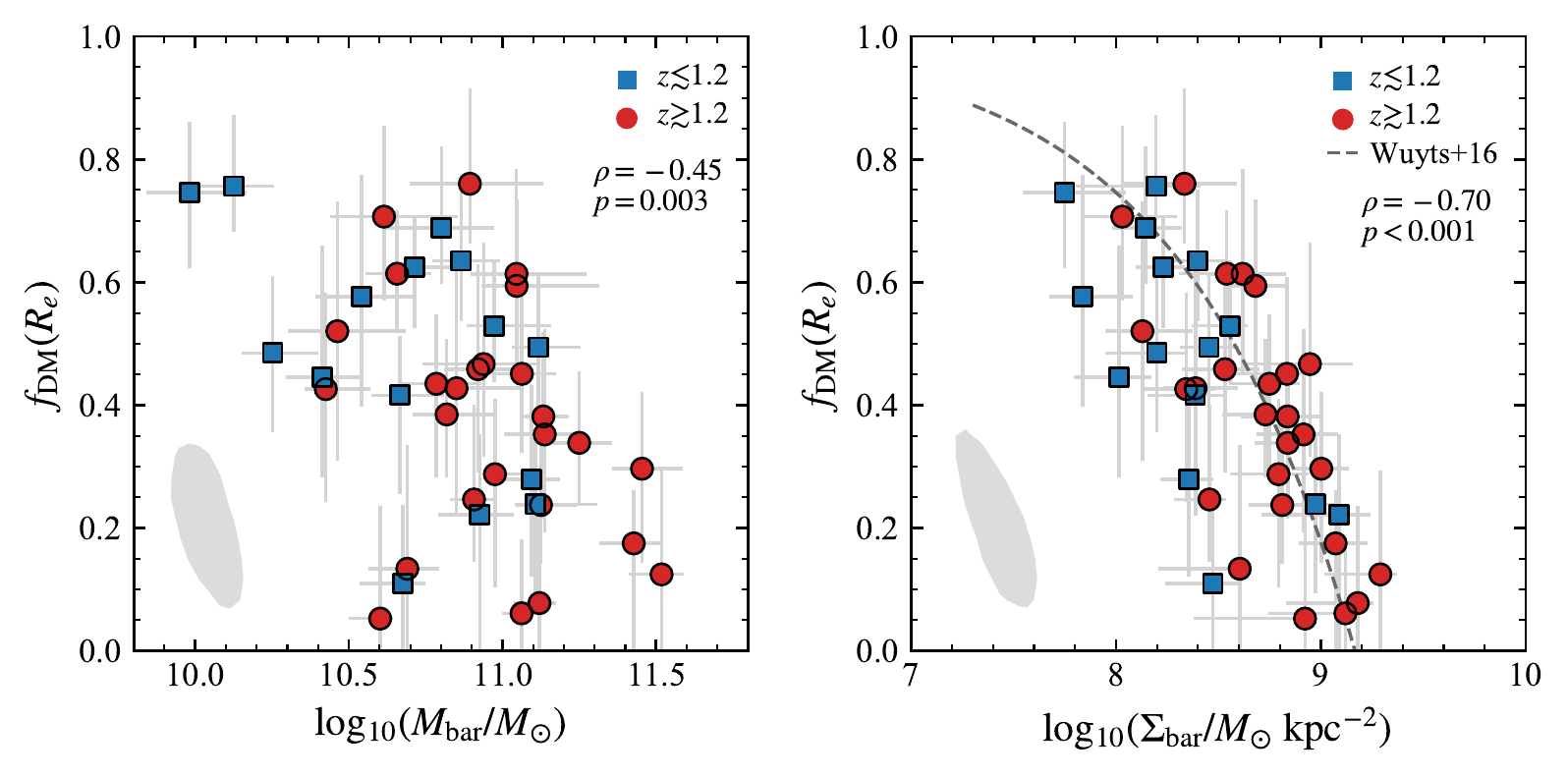}
\vglue -6pt
\caption{
Dark matter fraction versus baryonic mass (left) and versus baryonic surface density (right) 
using the best-fit results of the 1D MCMC analysis in this paper. 
The symbol definitions are the same as in Figure~\ref{fig:1D_chisq_vs_mcmc}. 
The median $1\sigma$ uncertainty contours are shown with filled grey regions. 
The grey dashed line in the right panel shows the relation between \Sigbar and \fDM from \citet{Wuyts16}. 
Spearman correlation coefficients and \textit{p}-values are also shown. 
We find very similar anti-correlations between \fDM and \lMbar (moderate, $2.7\sigma$) 
and between \fDM and \lSigbar (strong, $5\sigma$) as reported in \citetalias{Genzel20} 
(their Figure~7, lower left and upper right, respectively), though the MCMC MAP \fDM values are 
a bit higher on average, particularly for some of the $z\sim2$ galaxies. 
}
\vglue 12pt
\label{fig:1D_mcmc_fDM_lMbar_lSigbar}
\end{figure*}

% ++++++++++++++++++++++++++++

\begin{deluxetable*}{l c c c c l }[!ht]
\renewcommand{\arraystretch}{1.035} 
\setlength{\tabcolsep}{0.1in} 
\tabletypesize{\scriptsize}\tablenum{4}
\tablecaption{Priors for 2D MCMC Fitting\tablenotemark{\tiny{a},\tiny{b}}}
\label{tab:2d_mcmc_fit_priors}
\tablehead{
\colhead{ID} & \colhead{\lMbar} & \colhead{\fDM} & \colhead{\sigmaint} & 
\colhead{\redisk} & \colhead{$\Vsys$}\\[-4pt]
\colhead{ } & \colhead{[dex]} & \colhead{---} & \colhead{[$\mathrm{km\ s^{-1}}$]} & 
\colhead{[kpc]} & \colhead{[$\mathrm{km\ s^{-1}}$]} 
} 
\startdata
zC\_403741 & 		Gaus$(10.86, 0.2)$; $[9, 13]$ & Flat$[0,1]$ & Flat$[5,300]$ & \textit{Fixed} & Flat$[-100,100]$  \\
D3a\_6397 & 		Gaus$(11.34, 0.2)$; $[9, 13]$ & Flat$[0,1]$ & Flat$[5,300]$ & \textit{Fixed} & Flat$[-130,70]$  \\
zC\_405501 & 		Gaus$(10.43, 0.2)$; $[9, 13]$ & Flat$[0,1]$ & Flat$[5,300]$ & Gaus$(5.0, 2)$; $[1,15]$ & Flat$[-108,92]$  \\
BX389 & 			Gaus$(10.94, 0.2)$; $[9, 13]$ & Flat$[0,1]$ & Flat$[5,300]$ & \textit{Fixed} & Flat$[-180,20]$  \\
zC\_407302 & 		Gaus$(10.81, 0.2)$; $[9, 13]$ & Flat$[0,1]$ & Flat$[5,300]$ & \textit{Fixed}  & Flat$[-150,50]$  
\\[2pt]
\arrayrulecolor{LightGrey}\hline
\arrayrulecolor{black}
zC\_406690& 		Gaus$(11.09, 0.2)$; $[9, 13]$ & Flat$[0,1]$ & Flat$[5,300]$ & \textit{Fixed}  & Flat$[-75,125]$  \\
BX610 & 			Gaus$(11.52, 0.2)$; $[9, 13]$ & Flat$[0,1]$ & Flat$[5,300]$ & Gaus$(4.9, 2)$; $[1,15]$ & Flat$[-100,100]$  \\
K20\_ID7 & 		Gaus$(10.76, 0.2)$; $[9, 13]$ & Flat$[0,1]$ & Flat$[5,300]$ & \textit{Fixed} & Flat$[-100,100]$  \\
K20\_ID6 & 		Gaus$(10.80, 0.2)$; $[9, 13]$ & Flat$[0,1]$ & Flat$[5,300]$ & Gaus$(5.0, 2)$; $[1,15]$ & Flat$[-100,100]$  \\
zC\_400569 & 		Gaus$(11.50, 0.2)$; $[9, 13]$ & Flat$[0,1]$ & Flat$[5,300]$ & \textit{Fixed} & Flat$[-100,100]$ 
\\[2pt]
\arrayrulecolor{LightGrey}\hline
\arrayrulecolor{black}
BX482 & 			Gaus$(11.00, 0.2)$; $[9, 13]$ & Flat$[0,1]$ & Flat$[5,300]$ & Gaus$(5.8, 2)$; $[1,15]$ & Flat$[-40,160]$  \\
D3a\_15504 & 		Gaus$(11.28, 0.2)$; $[9, 13]$ & Flat$[0,1]$ & Flat$[5,300]$ & \textit{Fixed} & Flat$[-100,100]$  \\
D3a\_6004 & 		Gaus$(11.70, 0.2)$; $[9, 13]$ & Flat$[0,1]$ & Flat$[5,300]$ & \textit{Fixed} & Flat$[-100,100]$  \\
GS4\_42930 & 		Gaus$(10.65, 0.2)$; $[9, 13]$ & Flat$[0,1]$ & Flat$[5,300]$ & \textit{Fixed} & Flat$[-100,100]$ \\[2pt]
\enddata
\tablenotetext{a}{Gaussian priors are noted as ``Gaus$(\mathrm{center}, \mathrm{stddev})$'', 
and are additionally bounded within the range $[\mathrm{lower},\mathrm{upper}]$. 
Flat priors and their boundaries are denoted by ``Flat$[\mathrm{lower},\mathrm{upper}]$''.} 
\vspace{-6pt} 
\tablenotetext{b}{When not fixed, the priors on \lMbar, \fDM, \sigmaint, and \redisk 
are the same for both the 1D and 2D MCMC fitting.}
\vspace{-10pt} 
\end{deluxetable*}

For the 1D fits, the MCMC chain sampling is run using 1000 walkers, with 200 steps after a burn-in of 50 steps. 
With these settings, the chains for all fits over all objects have a final mean acceptance fraction between 0.2 and 0.5. 
We then adopt the \textit{maximum a posteriori} (MAP) values of the fit parameters as the best-fit values, 
where we jointly analyze the posteriors for the free-parameter priors (to account for degeneracies in the posterior distributions). 
The lower and upper $1\sigma$ uncertainties are then determined from the shortest interval containing 68\% 
of the marginalized posterior for each parameter (as discussed in Appendix~\ref{sec:DYSMALfitting}). 
The best-fit MAP values for the 1D MCMC fitting are listed in Table~\ref{tab:results_NFW}.

\vspace{6pt}

% +++++++++++++++++++++++++++++++++++++++++++++++++++++++
\subsection{Fitting 2D Kinematic Maps}
\label{sec:model_fitting_2D}

For the 2D modeling of the sensitive, deepest-possible data (composite or AO-only), 
we also simultaneously fit the velocity and dispersion maps. 
We begin with the same model setup as for the 1D fits, but additionally allow the systemic velocity $\Vsys$ to vary. 
We fix the kinematic major axis position angle, inclination, and spatial center for the 2D fits, as 
asymmetries or non-circular motions in the 2D maps can impact these parameters. 
For the 2D fits, the MCMC sampling is also run with 1000 walkers, but the chains are run longer than for the 1D fitting. 
We adopt fiducial settings of 300 steps after a burn-in of 100 steps, but some objects require longer burn-in periods. 
For some objects, \redisk is poorly constrained, so the 2D fits are repeated while fixing \redisk 
to the best-fit values from the 1D MCMC fitting (given in Table~\ref{tab:results_NFW}). 
The priors used for the 2D fitting, and whether \redisk is fixed or free, are listed in Table~\ref{tab:2d_mcmc_fit_priors}.

\vspace{12pt}

%%%%%%%%%%%%%%%%%%%%%%%%%%%%%%%%%
%%%%%%%%%%%%%%%%%%%%%%%%%%%%%%%%%
\section{Comparison of 1D kinematic disk fitting using least-squares and MCMC}
\label{sec:compG20}

In Figure~\ref{fig:1D_chisq_vs_mcmc}, we compare the best-fit values of the dark matter fraction \fDM, 
baryonic mass \lMbar, intrinsic dispersion \sigmaint, and disk effective radius \redisk from \citetalias{Genzel20} and this analysis. 
We stress that both analyses are based on the same 1D observed rotation and dispersion profiles. 
Overall, we find fairly good agreement between the two fitting methodologies, 
with the MCMC results finding the same overall trends and results as in \citetalias{Genzel20}. 

The MCMC analysis tends to find slightly higher dark matter fractions, 
with a median offset of $\langle \Delta \fDM \rangle = \langle \fDM_{\mathrm{MCMC}} - 
\fDM_{\mathrm{least-sq.}} \rangle = 0.12$ between the analyses. 
We also find correspondingly slightly lower baryonic masses than the least-squares analysis, 
with $\langle \Delta \lMbar \rangle = -0.04\unit{dex}$. 
The derived velocity dispersions also tend to be higher in the MCMC, with offset 
$\langle \Delta \sigmaint \rangle = 8.64\unit{km\ s^{-1}}$. 
Finally, the disk effective radii are overall in very good agreement, though 
for smaller galaxies the best-fit radii are slightly higher in the MCMC analysis 
(with an overall median offset of $\langle \Delta \redisk \rangle = 0.65\unit{kpc}$). 
These median differences are within the typical uncertainties for the two analyses 
(median uncertainties of $\sim\!0.1{-}0.14$ for \fDM, $\sim\!0.1{-}0.13\unit{dex}$ for \lMbar, 
$\sim\!5{-}8\unit{km\ s^{-1}}$ for \sigmaint, and $\sim\!1{-}2\unit{kpc}$ for \redisk).

Generally, these differences in values reflect some of the inherent degeneracies in our models 
(as discussed in Appendix~\ref{sec:dysmal_degen_plp}), modulated by the adopted priors and constraints. 
For the MCMC analysis, looser constraints on both \sigmaint and \redisk are adopted, as well a Gaussian prior for \lMbar. 
In contrast, the least-squares analysis generally had tight constraints on \sigmaint (based on the outermost dispersion points), 
and adopted top-hat bounds for \Mbar. 
In particular, the looser constraints on \sigmaint impacted the MCMC fitting. 
The increased parameter flexibility resulted in cases where the likelihood 
is maximized by a higher \sigmaint, as the improved match of the model velocity curve to the data 
outweighed the increased discrepancy with the dispersion profile. 
In other cases, the overall observed dispersion profile resulted in higher likelihood for models with increased \sigmaint. 
This tendency of higher \sigmaint (from the looser \sigmaint prior) combines with the effect of the adopted Gaussian priors 
on \redisk and \lMbar to impact the best-fit results of our simultaneous 4-parameter MCMC fitting. 
The higher \sigmaint and marginally higher \redisk values tend to shift up the DYSMAL model $\vcirc(\re)$ degeneracy ellipse 
(already moving towards higher \fDM), and the slightly lower \lMbar values translate into higher \fDM values 
(see Figure~\ref{fig:fdm_mbar_degen_illustration}, Appendix~\ref{sec:dysmal_degen_plp}).

Despite these small differences in the recovered values between these two approaches, 
the MCMC analysis confirms that  
\textbf{the majority of galaxies in our sample (\hbox{$\boldsymbol{\sim\!70\%}$}), particularly at $\boldsymbol{z\gtrsim1.2}$, 
are baryon-dominated on galaxy scales, with $\boldsymbol{\fDM\lesssim0.5}$.} 
For $\sim\!\!30\%$, the dark matter fractions are similar to or less than that of ``maximal disks'' 
($f_{\mathrm{DM,max}} < 0.28$; \citealt{Courteau15}). 
Moreover, the MCMC results recover the same overall trends seen in \citetalias{Genzel20}, such as 
the strong anti-correlation of dark matter fraction with baryonic surface density 
and the moderate (and less significant) anti-correlation of dark matter fraction with total baryonic mass, 
which we show in Figure~\ref{fig:1D_mcmc_fDM_lMbar_lSigbar}. 
\textbf{Overall, we find that the least-squares and MCMC results are in good agreement.}

%% ++++++++++++++++++++++++++++++++++++++++++++++++++++++++++++++
% ++++++++++++++++++++++++++++
\begin{figure*}[ht!]
\vglue -4pt
\centering
\includegraphics[width=0.75\textwidth]{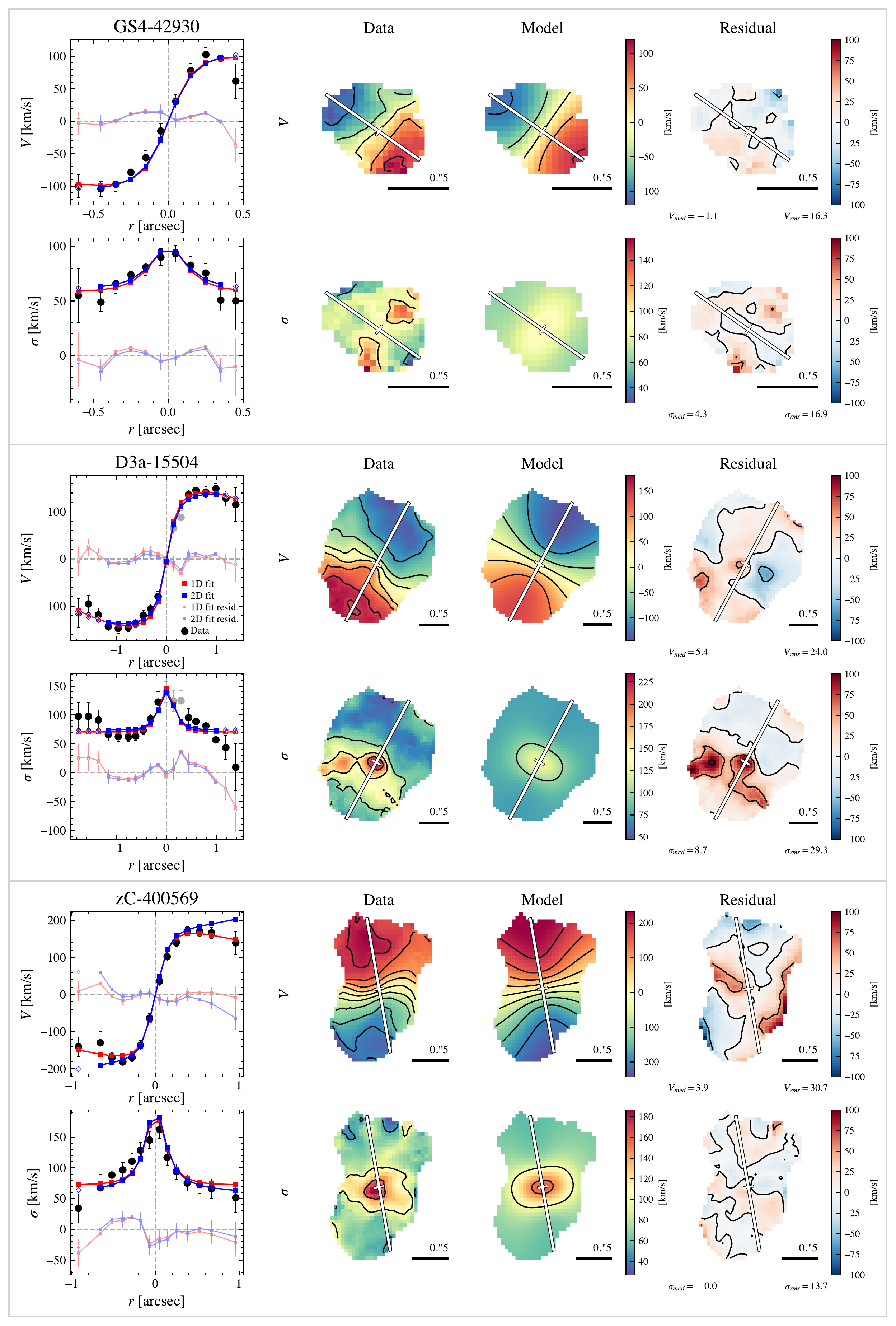}
\vglue -5pt
\caption{
Example 2D fits for 3 galaxies in the RC41 sample. 
The composite 2D velocity (top) and dispersion (bottom) maps are shown in the second column for each galaxy. 
The best-fit models and residuals from the 2D fitting are shown in the third and fourth columns. 
The major axes and centers are marked with white lines on the 2D maps, 
and contours are marked in $50\unit{km\ s^{-1}}$ intervals. 
In the first column, we compare the 1D observed profiles (black circles) with the 1D best-fit models (red squares) 
and the 2D best-fit models, extracted within the same 1D apertures/PV diagrams 
(blue points, with filled squares and open diamonds showing 
\hbox{$\leq\!R_{\mathrm{out,2D}}$} and \hbox{$>\!R_{\mathrm{out,2D}}$}, 
The 1D residuals for the 1D and 2D models are shown in light red and blue, respectively. 
Overall, the 1D and 2D fits of the disk kinematics 
show fairly good agreement, but there are deviations towards larger radii 
that likely reflect effects from non-circular features. 
The axisymmetric disk models used in this study work well for galaxies with fairly symmetric kinematics (e.g., GS4-42930). 
Non-circularities such as PA twists can lead to differences between the 1D and 2D fits (e.g., zC-400569), 
but in other cases both fits agree despite such features (e.g., D3a-15504). 
Previously, the 2D disk modeling of D3a-15504 (and other subsample objects) 
by \citealt{Cresci09} also revealed the primary disk rotation and secondary residual features in this galaxy.
}
\label{fig:example_2D_fits}
\end{figure*}
% ++++++++++++++++++++++++++++
%% ++++++++++++++++++++++++++++++++++++++++++++++++++++++++++++++

% ++++++++++++++++++++++++++++++++++++++++++++++++++++++++++++++
% ++++++++++++++++++++++++++++++++++++++++++++++++++++++++++++++
\section{Comparison of Disk Fitting Using 1D and 2D Kinematics}
\label{sec:comp1D2D}

For the RC41 data set, we performed our dynamical analysis using extracted 1D kinematic profiles, 
to maximize the data S/N and to push to the largest possible radii. 
For axisymmetric distributions, 
2D maps of the galaxy velocity and dispersion fields (or additional higher-order moments) 
encode additional information about the kinematic position angle and galaxy inclination, and can provide 
additional constraints on the disk effective radius, \redisk, or the bulge-to-total ratio, \bt.
Given sufficient 2D S/N, it would thus be possible to independently fit for these parameters, 
instead of adopting values derived from imaging under the assumption that they also apply 
to the underlying mass distribution, which may be incorrect given intrinsic mass-to-light gradients or 
optically thick dusty regions in the centers of galaxies. 
It should be noted, however, that 2D maps can reveal noncircular motions, which capture other physical processes 
superimposed on the regular disk rotation.

Here we compare 1D and 2D fitting results for a subset of 14 galaxies in the RC41 sample, 
in order to examine how much dynamical information is captured along the major axis, 
and to consider the relative benefits of fitting in 1D versus 2D. 
We then briefly examine non-circular motions seen in the 2D residual maps for one galaxy, 
and present an example toy model that could describe this additional kinematic signature.

% ++++++++++++++++++++++++++++++++++++++++++++++++++++++++++++++
\subsection{Disk Modeling: Kinematics Well-Measured by 1D Fitting}
\label{sec:comp1D2D_diskfit}

In order to maximize the depth of the 2D maps (and therefore pushing to the maximum outer radii), we perform 
fitting using 2D maps derived from the deepest possible data --- generally 
the composite cubes combining all available AO-assisted and seeing-limited data. 
The measurement of the 2D maps and the 2D fitting methodology are presented 
in detail in Sections~\ref{sec:data} and \ref{sec:model_fitting_2D}. 
The key difference between the 1D and 2D fitting is in the treatment of \redisk:
it is a free parameter for the 1D modeling, but for many cases is poorly constrained by the 2D maps.
Thus, for most 2D fits we fix \redisk to the best-fit 1D MCMC value. 
For the 2D fitting we additionally fix the kinematic center and orientation (\PA, inclination) for each galaxy, 
as these parameters can be strongly impacted by asymmetries or non-circular kinematic 
features (as discussed below).

As an example of the 2D versus 1D fits, in Figure~\ref{fig:example_2D_fits} 
we show the 2D maps, best-fit models, and residuals for three galaxies (second through fourth columns). 
For comparison with the 1D fits, we also show the best-fit 2D models extracted along the major axis 
using pseudo-slits versus the observed and best-fit 1D profiles (first column). 
There is relatively good agreement in the 1D profiles between the data, 1D and 2D best-fit models, 
though the 1D profiles from the 2D models diverge from the data towards the outskirts in some cases. 
As discussed below, we attribute these deviations 
primarily to the impact of non-circular kinematics or asymmetric features on the 2D fits.

When comparing the 1D and 2D fits, we find a relatively good agreement between the two measurements of \fDM overall. 
However, on average the 2D fits tend to find larger dark matter fractions. 
Furthermore, a good fraction of the galaxies show discrepancies between 
$\fDM_{\rm 1D}$ and $\fDM_{\rm 2D}$, up to differences of \hbox{$|\Delta\fDM|\sim0.45$}. 
For the baryonic masses, there is excellent agreement above $\lMbar\sim11$, 
but we tend to find a  \hbox{$\sim\!-0.2\unit{dex}$} offset between the 2D and 1D values 
for the lower masses (and accordingly generally find higher 2D dark matter fractions). 
There is scatter between the measured 1D and 2D intrinsic dispersion values, 
as \sigmaint is sensitive to off major-axis features and deviations from pure rotational motion in the 2D maps. 
Two factors could be responsible for these discrepancies.

First, there are differences in the 1D and 2D data maximal radial extent (i.e., \Rout, as noted in Figure~\ref{fig:example_2D_fits}). 
The apertures used to extract the 1D profiles help to optimize the data S/N, pushing the 
kinematic profiles out to large radii, particularly when flared apertures are used. 
In contrast, the 2D maps are limited by the per-spaxel S/N, and thus are generally limited to smaller galactic radii. 
For most of the objects in this subset, $R_{\mathrm{out,1D}}$ is greater than $R_{\mathrm{out,2D}}$ by $\sim0.5{-}3\unit{kpc}$. 
The extra radial coverage of the 1D profiles can help to break degeneracies between model parameters, 
as this better probes regions where dark matter is expected to become more important relative to the baryons. 
The more extended 1D profiles can also better constrain 
the intrinsic velocity dispersion, as \sigmaint is best probed in regions less impacted by beam smearing (i.e. large galactic radii). 
We test the relative importance of the data radial extent in our fits by repeating the 
1D MCMC fitting using profiles truncated to more closely match $R_{\mathrm{out,2D}}$ (e.g., see 
left panels of Figure~\ref{fig:example_2D_fits}). 
These fits find nearly identical results within the uncertainties to the untruncated 1D fits (with the exception 
of \sigmaint for zC-407302, where the extended 1D profile was crucial to constraining the intrinsic dispersion). 
Thus, we conclude that while for our subset of 14 galaxies 
radial extent can play a role in constraining dynamical components, 
it is not the primary driver of differences between the 1D and 2D fit results.

Second, and more importantly, non-circular kinematics or asymmetric features 
impact our kinematic fitting, particularly in 2D given the extra off-major axis information. 
As we fit the data with axisymmetric mass and kinematic models, any asymmetry 
or non-circular motions can pose problems to the recovery of the intrinsic underlying circular motion of the galaxies. 
The major axis 1D profiles should be much less sensitive to such issues, 
as many of these features are not captured along major axis kinematic cuts. 
We find that most of the objects in our 2D analysis exhibit kinematic asymmetries or non-circular features, 
which, depending on their nature and strength, can significantly affect results of modeling with axisymmetric-only models. 
For example, in Figure~\ref{fig:example_2D_fits}, we show two galaxies 
exhibiting PA twists in the 2D maps. For one object (zC-400569; bottom), this PA twist is likely responsible for 
the difference in the 1D and 2D fits, while for the other (D3a-15504; middle) the twist does not strongly impact the 2D fit. 
In contrast, we would expect more similar 1D and 2D fits for galaxies with fairly symmetric 2D kinematics
(e.g., GS4-42930; Figure~\ref{fig:example_2D_fits}, top). 
By probing only the major-axis kinematics, the 1D fits may better reflect the 
underlying disk kinematics and mass distributions than the 2D fits. 
The non-circular or asymmetric features may drive the 2D fit results away from the intrinsic galaxy properties, 
complicating a direct comparison between the 1D and 2D values. 
However, from our current small 2D sample, it is difficult to determine what type and strength of 
non-circular or asymmetric features impact the 2D disk kinematic fitting.

Based on the general agreement between the 1D and 2D fit \fDM values 
in the modeling framework we consider (i.e., oblate disk, bulge, and halo), we conclude that 
\textbf{the primary kinematics of a rotating galaxy are well constrained using 1D fitting  along the major axis only.} 
Given the challenges to the 2D kinematic modeling (i.e., data extent; purely axisymmetric models), 
the underlying disk kinematics may be better constrained by fitting in 1D than in 2D.

%% ++++++++++++++++++++++++++++++++++++++++++++++++++++++++++++++
% ++++++++++++++++++++++++++++
\begin{figure*}[t!]
\vglue -4pt
\centering
\includegraphics[width=0.85\textwidth]{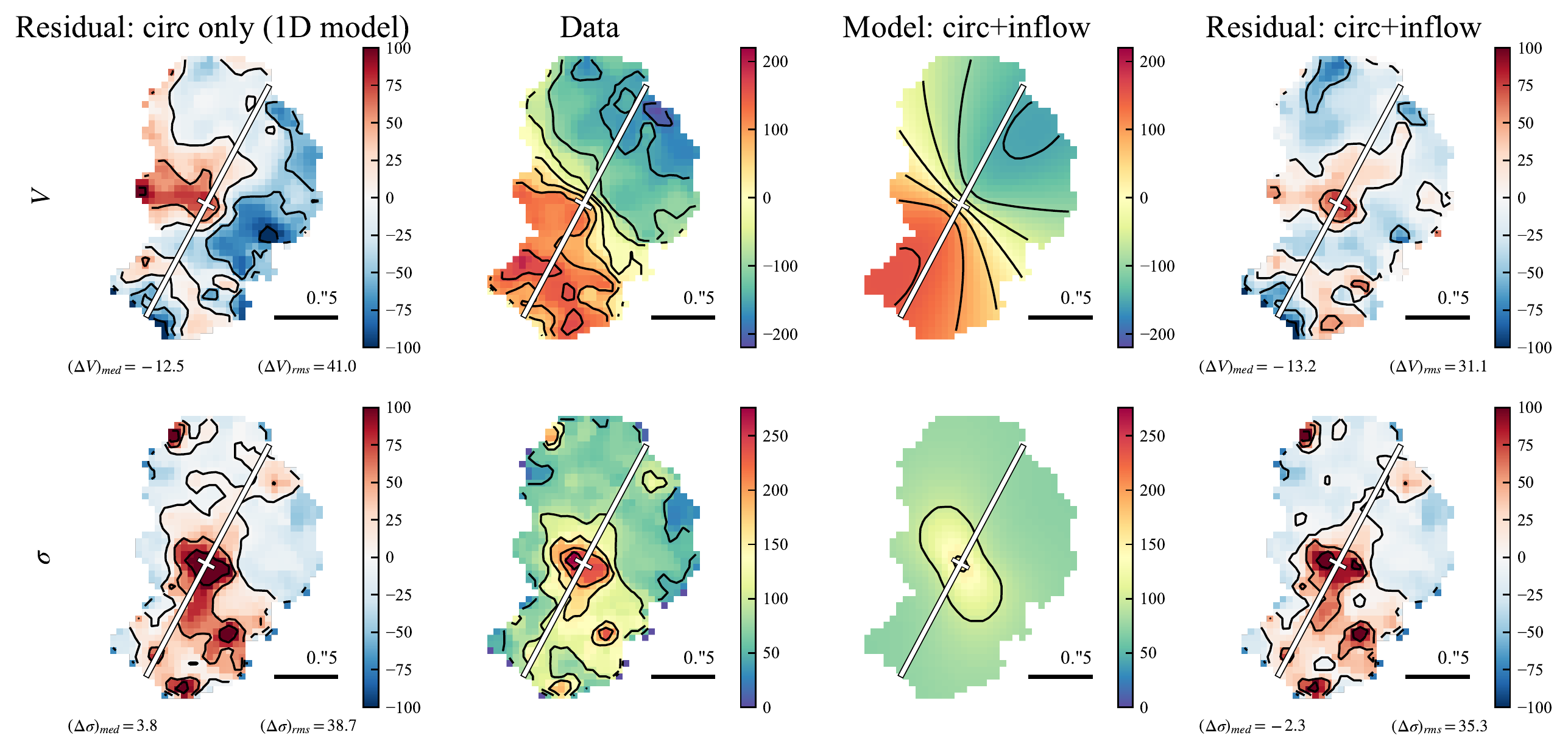}
\vglue -6pt
\caption{
Velocity (top) and dispersion (bottom) maps (with scales in units of $\mathrm{km\ s^{-1}}$) of one object, D3a-15504, 
from the 2D fitting subset that exhibits non-circular motions.
Here we use maps derived using only the AO data, to maximize the spatial resolution in the central regions of the galaxy. 
The maps have been median smoothed with a kernel of $3\times 3$ pixels, 
and we show contours every $50\unit{km\ s^{-1}}$. 
The first column shows the residual maps for the galaxy+halo model determined with 1D MCMC fitting, 
with appropriate $\xcenter, \ycenter, \Vsys$ applied.  
The observed AO maps are shown in the second column. 
We then show an example 2D model in the third column, where we use the same disk+bulge and halo parameters 
from the 1D fit but manually add a constant radial inflow velocity of $v_{r}\sim90\unit{km\ s^{-1}}$ (determined by eye after a grid search). 
The residuals from this composite rotation and radial motion model are shown in the last column. 
While this is \emph{not} a fit to the data, this example rotation+inflow model reduces 
the bimodal velocity residual at the center of the galaxy that is perpendicular to the major axis. 
The test composite model also largely captures the large-scale ``S'' twist in the rotation velocity. 
The AGN-driven outflow in this galaxy likely explains the remaining minor-axis velocity bimodality and central dispersion peak.
}
\label{fig:example_radial_2D}
\end{figure*}
% ++++++++++++++++++++++++++++
%% ++++++++++++++++++++++++++++++++++++++++++++++++++++++++++++++

% ++++++++++++++++++++++++++++++++++++++++++++++++++++++++++++++
\subsection{Other Kinematic Signatures in 2D Maps}
\label{sec:comp1D2D_noncirc}

While 1D major axis profiles capture the key rotational kinematics of disk galaxies, 
2D (or 3D) kinematic data are needed to probe higher-order dynamical effects caused by, 
e.g., perturbations, radial inflows, or outflows. 
Most of the galaxies in our 2D subsample exhibit signatures of noncircular motions, 
highlighted in previous work (for example, \citealt{Genzel06,Genzel08,Genzel11,Genzel17,Genzel20}, \citealt{ForsterSchreiber09,ForsterSchreiber18}, \citealt{Cresci09}). 
These features include, notably, twists in the kinematic major axis, 
other residual features along the minor axis, perturbations likely stemming from interactions with a 
neighbor/lower-mass satellite, and features resulting from nuclear outflows due to AGN or 
stellar feedback from bright off-center clumps. 
Some objects show signs of multiple types of features.

To illustrate the secondary signatures from noncircular motions seen in 
the 2D kinematics, we examine the case of D3a-15504, 
for which the kinematics show some of the most suggestive features of radial inflow \citep{Genzel06}. 
We construct a toy model combining disk rotation with radial inflow as follows. 
We first subtract the best-fit galaxy+halo model (determined from the 1D MCMC fits) 
from the 2D AO kinematic maps, yielding initial velocity and dispersion residual maps. 
For this galaxy, the velocity residual (Figure~\ref{fig:example_radial_2D}, top left) 
shows a central residual bimodality, and a larger-scale bimodal ``twist''. 
The dispersion residual shows a large central excess, 
which primarily reflects the nuclear outflow seen in this galaxy (\citealt{Genzel06}, \citealt{ForsterSchreiber14}). 
We then construct a second model (shown in Figure~\ref{fig:example_radial_2D}, third column), 
superimposing a constant radial inflow component on top of the best-fit 1D MCMC galaxy+halo model. 
The radial inflow velocity $v_{r}$ is manually adjusted (explored with a grid search). 
We find that a constant radial inflow component of $v_{r}\sim90\unit{km\ s^{-1}}$
can explain much of the original kinematic residuals (see Figure~\ref{fig:example_radial_2D}, last column), 
removing much of the large-scale ``S'' twist from NE to SW and 
the central velocity bimodality seen in the galaxy+halo-only velocity residual map. 
The remaining minor-axis velocity bimodality and central dispersion peak are likely due to the 
AGN-driven outflow in this galaxy (as noted in \citealt{Genzel06}), 
but we do not attempt to model the outflow for this toy model. 
While including an inflow component is not necessary to 
improve the 2D disk fit properties for D3a-15504 (as the 1D and 2D fits agreed well, see Figure~\ref{fig:example_2D_fits}), 
for other objects the inclusion of secondary kinematic components could lead to refined 2D disk fits 
that are in better agreement with the 1D major axis fits.

This initial exploration highlights how deep 2D kinematic maps, particularly from 
high-resolution adaptive optics-assisted observations, can constrain other dynamical signatures beyond just disk rotation. 
An extended analysis of the non-circular kinematic signatures in this sample (and in further observations) 
will be explored in greater detail in future work.

%%%%%%%%%%%%%%%%%%%%%%%%%%%%%%%%%
%%%%%%%%%%%%%%%%%%%%%%%%%%%%%%%%%
\section{Discussion}
\label{sec:disc}

\subsection{Multiple Studies and Approaches Find Consensus: Massive SFGs at $z\sim1{-}2$ are Baryon-Dominated}
\label{sec:disc_consistency}

In the previous sections, we have shown that multiple fitting approaches produce similar results 
for our sample of 41 massive, extended galaxies at $z\sim1-2$. 
Whether performing least-squares fitting (as in \citetalias{Genzel20}) or MCMC sampling of 1D rotation curves 
extracted along the major axis, or if using the more detailed 2D rotation and dispersion maps, 
the best-fit values of \fDM, \lMbar, \sigmaint, and \redisk are all in relatively good agreement. 
These results are also consistent with those of \citet{Genzel17} and \citet{Ubler18}, 
which together analyzed seven galaxies that are included in the RC41 sample.

One key result from our analyses of the RC41 sample is that there is a strong anti-correlation between \fDM and the 
baryonic surface density (Figure~8 of \citetalias{Genzel20}; also Figure~\ref{fig:1D_mcmc_fDM_lMbar_lSigbar}). 
However, we selected large galaxies to maximize the number of spatial resolution elements, 
and thus our sample does not include many small, high-density galaxies. 
We thus compare the RC41 results to  the findings of ``inner dynamics'' analyses, 
which use kinematic signatures on less-extended scales (i.e., $\lesssim\re$) 
in combination with separate measurements of \Mbar or \Mstar to estimate the baryonic and dark matter fractions. 
Kinematic measurements on these scales can be performed 
for larger samples extending to smaller galaxy sizes than when fitting to outer rotation curves, 
as the depth and resolution requirements are lower.

There is good agreement between observations from such ``inner dynamics'' studies of 
individual massive star-forming galaxies and our RC41 analysis. 
\citet{Wuyts16} find a very similar trend between \fDM and \lSigbar for a sample of 240 galaxies (as discussed in \citetalias{Genzel20}),  
and other studies find high baryon fractions towards smaller sizes and higher densities, and with increasing redshift 
(e.g., \citealt{vanDokkum15}, \citealt{Price16,Price20}). 
These results are also consistent with the simulation work of galaxies at this epoch. 
In particular, \citet{Lovell18} find a similar anti-correlation of \fDM and baryonic surface density 
(as shown in \citetalias{Genzel20}; M. R. Lovell 2020, private communication), 
and \citet{Zolotov15} find that galaxies are baryon-dominated during times when they are very compact. 
The general trend of decreasing \fDM towards higher redshifts is 
further supported by the stellar and baryonic Tully-Fisher zero-point evolution found by \citet{Ubler17} 
and the numerical simulations presented by \citet{Teklu18}.

The literature studies discussed above all constrain the enclosed \fDMg on galaxy scales (i.e., $r=\re$ or \redisk) 
using the observed kinematics of individual galaxies. 
Other studies stack the rotation profiles of many galaxies to probe out to even larger galactic radii. 
\citet{Lang17} stacked the RCs of rotating disk galaxies at $z\sim0.6{-}2.6$, 
with the individual curves normalized at the peak $\vrot$ and corresponding radius, 
and found an average dropping outer RC. 
By comparing the stack to models consisting of thick exponential disks with \sigmaint typical of the stacked sample 
embedded in a NFW halo, \citet{Lang17} found that high baryonic fractions on galactic scales are required 
to match the stacked RC, in excellent agreement with the findings of \citetalias{Genzel20} and this analysis. 
A different approach was adopted by \citet{Tiley19}, who instead stacked RCs normalized at $3\rd\sim1.8\re$ 
based purely on the stellar light distribution. 
\citet{Tiley19} then compared their stacked RCs to models 
combining an infinitely thin exponential disk and a pseudo-isothermal dark matter halo, 
and found the stacks are consistent with low average baryonic fractions within $6\rd\sim3.6\re$ (i.e., high average \fDMg). 
However, it is difficult to compare the results between these studies because of the different methodologies, 
and importantly, the sample selections differ and thus differently represent the underlying population. 
We stress also the importance of the radius within which the enclosed \fDMg is referenced, 
as it obviously will increase towards outer regions. 
As highlighted by \citet{Ubler18}, and from examination of the results for the RC41 sample, 
large or even dominant enclosed \fDMg around \hbox{$\sim3.6\re$} is not in contradiction with 
strongly sub-dominant \fDMg at $\lesssim1\re$. 
More importantly, the enclosed $\fDMg(r)$ may provide clues as to the relative concentration of baryons and dark matter 
as galaxies build up over time, as discussed in \citetalias{Genzel20}.

% ++++++++++++++++++++++++++++
\begin{figure*}[t!]
\centering
\vglue -6pt 
\hglue -4pt
\includegraphics[width=0.9\textwidth]{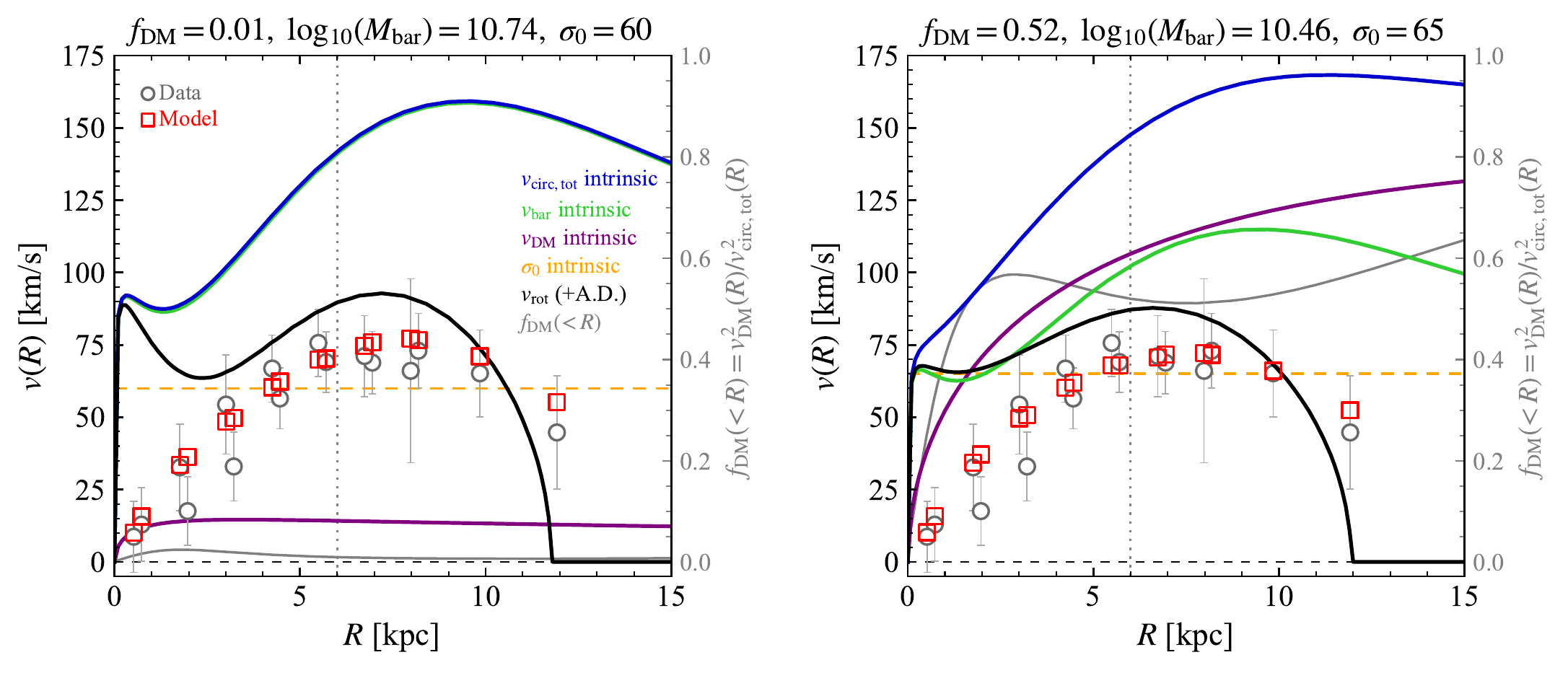}
\vglue -12pt
\caption{
Two example galaxy + halo models in an extreme case 
demonstrating the degeneracy of models with negligible ($\fDM=0.01$, left) 
and relatively high ($\fDM=0.52$, right) dark matter fractions because of the strong effects of 
asymmetric drift at relatively high intrinsic velocity dispersions ($\sigmaint\sim60{-}65\unit{km\ s^{-1}}$). 
The intrinsic baryonic, dark matter, and total circular velocity profiles are shown as green, purple, and blue solid lines, respectively, 
and the intrinsic velocity dispersion is marked with the dashed orange line. 
The intrinsic rotation velocity profile, including the effects of asymmetric drift (using Eq.~\ref{eq:asymm_drift}, as in \citealt{Burkert10}), 
is shown as the solid black line. 
The solid grey line shows the dark matter fraction. \redisk is marked with the dotted vertical line. 
The folded 1D observed velocity profile for zC-405501 is shown with the grey-outline points. 
We also extract the models from the fully forward-modeled cubes using the same apertures (red open squares). 
The respective \fDM and \lMbar of the models result in similar values of $\vcirc(\re)$, 
while the slight \sigmaint variation controls the disk truncation radius. 
Thus, depending on how large \sigmaint is and how well it is constrained, 
it is possible for asymmetric drift to contribute to the degeneracy between disk and halo, even at this early epoch ($z\sim1{-}2$) 
with less concentrated halos when the disk-halo conspiracy should be less problematic.
}
\vspace{4pt} 
\label{fig:vrot_int_example}
\end{figure*}

After accounting for analysis differences, these other studies are in remarkable agreement with the results of our RC41 work. 
This reinforces the growing consensus that massive star-forming galaxies at $z\sim1{-}2$ tend to be baryon-rich 
on galaxy scales (i.e., \hbox{$\sim\!1\re$}), with decreasing dark matter fractions as galaxies become more compact. 
Furthermore, these findings suggest that galaxies with low dark matter fractions have cored halos, 
as the virial masses inferred from the measured \fDM for a NFW profile fall far short of the predictions 
from stellar mass-halo mass relations (as shown in Figs.~9 \& 10 of \citetalias{Genzel20}). 
Theoretical studies, though primarily targeting dwarf galaxies or galaxy clusters, 
show that dark matter core formation can occur due to dynamical friction from infalling clumps (e.g., \citealt{ElZant01}, 
\citealt{Mo04}, \citealt{Johansson09}, \citealt{Romano-Diaz09}, \citealt{Goerdt10}, \citealt{Cole11}, \citealt{Nipoti15}), 
bars (e.g., \citealt{Weinberg07}), vigorous, fluctuating outflows (e.g., \citealt{Dekel86}, \citealt{Navarro96}, 
\citealt{Read05}, \citealt{Mashchenko08}, \citealt{Pontzen12,Pontzen14}, \citealt{Martizzi13}, \citealt{Freundlich20}, 
K. Dolag et al., in prep.), or impulsive heating from minor mergers (\citealt{Orkney21}). 
Similar physical processes could potentially lead to cored dark matter halos in 
the massive, gas-rich, star-forming galaxies at $z\sim1{-}2$.

%%%%%%%%%
\subsection{On the role of high velocity dispersion at high redshifts}
\label{sec:AD_highsig}

A complication of the measurement of dark matter fractions is the degeneracy between baryonic disk and dark matter halo. 
In local galaxies, the relatively high halo concentration contributes to the ``disk-halo'' conspiracy, 
where it is difficult to detangle the velocity profiles of the disk and halo on the galaxy scale, 
making it hard to distinguish between their mass contributions. 
At higher redshifts, lower concentrations of halos at fixed stellar mass should help to 
break this degeneracy, by moving the halo profile signatures to larger radii (as discussed in \citetalias{Genzel20}).

For galaxies with relatively high intrinsic velocity dispersion (e.g., $\sigmaint\gtrsim50{-}60\unit{km\ s^{-1}}$), 
the exponential disk asymmetric drift correction (from \citealt{Burkert10}) 
predicts a marked reduction of the rotation velocity at $\gtrsim\redisk$ 
and a truncation of the disk by roughly a few times the effective radius.  
Under the right conditions, and together with the impact of beam-smearing, 
this can result in very similar rotation curves for models with both low and high dark matter fractions. 
This added degeneracy complication highlights the importance of simultaneously 
fitting both the rotation and velocity dispersion profiles, 
to obtain the best possible constraints on the dynamical parameters.

We highlight an extreme example of the added complication of 
asymmetric drift to the disk-halo degeneracy in Figure~\ref{fig:vrot_int_example}: zC-405501. 
This galaxy has a very broad likelihood degeneracy between \fDM--\lMbar that is strongly impacted by the asymmetric drift correction. 
In the left panel, we show a low dark matter fraction model, with $\fDM=0.01$, $\lMbar=10.74$, and $\sigmaint=60\unit{km\ s^{-1}}$, 
while the right panel shows a moderately high dark matter fraction model of 
$\fDM=0.52$, $\lMbar=10.46$, and $\sigmaint=65\unit{km\ s^{-1}}$. 
All other parameters are the same between both models (i.e., $\bt=0.07$, $\redisk=6\unit{kpc}$, $\nSdisk=0.2$). 
We show the intrinsic baryon, dark matter, and total circular velocity curves for both cases (with the 
green, purple, and blue solid curves, respectively), and also mark \sigmaint (orange dashed line) 
and the dark matter fraction as a function of radius (grey solid line). 
The rotation velocity profile determined by applying the asymmetric drift correction to the total circular velocity 
is shown as the solid black line. 
The disparity of dark matter fractions is clearly seen in the halo velocity curves, 
but both models have a fairly similar total circular velocity at \redisk (dotted grey line). 
Furthermore, the asymmetric drift correction produces \vrot curves with similar shapes outside of the smallest radii, 
including similar truncation radii. 
We note that the slightly higher intrinsic dispersion of the high \fDM case 
($\sigmaint=65\unit{km\ s^{-1}}$ versus $60\unit{km\ s^{-1}}$) 
is responsible for matching the truncation radius of the lower \fDM case 
(as the truncation radius would be higher if \sigmaint were the same).

When including all observational effects in the models for this extreme case 
(inclination, beam smearing, and extraction in flared rectangular apertures; red open squares), 
both low and high dark matter fraction models describe the observed data (grey-outlined circles) fairly well. 
In particular, the intrinsic small radii differences have been washed out by beam smearing, and both cases 
follow the dropping profile of the observed velocity curve, thanks to the strong effects of the asymmetric drift correction. 
If \sigmaint is very well constrained from the dispersion profile, the added impact of asymmetric drift can be partially mitigated, 
but for the case of zC-405501, the uncertainties on $\sigma(r)$ (typically \hbox{$\sim10{-}15\unit{km\ s^{-1}}$}) 
allow room for a slightly higher \sigmaint to better match the falloff for the 52\% dark matter fraction case. 
Observations with higher spatial resolution can also help to break this added degeneracy, 
as the intrinsic rotation curves (solid black lines) exhibit shape differences at small radii.

Nonetheless, the similarity of the 1D ``observed'' model profiles in these two cases highlights 
how it is still possible to have strong disk-halo degeneracies, even at these redshifts, 
as in the high-dispersion limit (with some uncertainty) the asymmetric drift correction 
can produce strong turnovers even in dark-matter dominated models. 
We note that the pressure support corrections used here are based on simple assumptions (i.e., a self-gravitating disk). 
Further constraints will help to determine the most applicable pressure correction 
for galaxies at these redshifts.

%%%%%%%%%%%%%%%%%%%%%%%%%%%%%%%%%%%%%%%%%%%%%%%%%%%
%%%%%%%%%%%%%
\section{Summary}
\label{sec:summary}

In this paper, we present a companion analysis of the kinematics for a sample of 41 large, massive, star-forming galaxies 
at $z\sim1{-}2$, which were first published in \citet{Genzel20} (\citetalias{Genzel20}). 
As in \citetalias{Genzel20}, we fit the 1D rotation curves using fully forward-modeled 3D kinematic models 
extracted to match the observations, but for this analysis we use MCMC sampling to derive 
maximum a posterior (MAP) ``best-fit'' values and to estimate the fit uncertainties. 
We additionally fit the kinematics for a subset of 14 galaxies using the spatially-resolved 2D rotation and dispersion maps. 
Our key findings are as follows: 
%%%%
\begin{enumerate}[label=(\roman*),leftmargin=16pt, itemsep=4pt,topsep=6pt, partopsep=4pt,parsep=2pt]
\item \textbf{Multiple approaches reinforce the finding that massive SFGs at $\boldsymbol{z\sim1{-}2}$ 
are baryon-dominated on galactic scales.}
We find good agreement between the measured kinematic and mass parameters for the 
1D least-squares analysis of \citetalias{Genzel20} and the 1D MCMC analysis presented here. 
Additionally, there is relatively good agreement between the \fDM values measured from 
the 1D and 2D MCMC fitting analysis for the subset of galaxies considered. 
The agreement between these three fitting methodologies, and with the results from other work 
(e.g., \citealt{vanDokkum15}, \citealt{Wuyts16}, \citealt{Price16, Price20}), 
demonstrates the robustness of our findings that massive galaxies at these redshifts
are generally baryon-dominated, with low galaxy-scale ($\sim1\re$) dark matter fractions. 
\item  \textbf{Primary disk kinematics are well captured along the major axis.} 
The agreement of the 1D and 2D kinematic fitting further supports 
that most of the kinematic information about disk dynamics and intrinsic dispersion for 
these high-redshift galaxies is encoded along the major axis. 
\item \textbf{Evidence for noncircular motions.} 
The data show evidence for noncircular motions, in addition to the 
disk kinematics for many of the galaxies in our 2D analysis subset. 
These features can affect the results of modeling in 2D. 
We show that a toy model with constant radial inflow can explain some of 
the noncircular residuals for one of these objects (D3a-15504).
\end{enumerate}
%%%%

Future work will expand on the analysis presented here and in \citetalias{Genzel20}. 
Efforts are ongoing to construct even larger samples of high-quality individual rotation curves, 
that will provide further insights into population and evolutionary trends in the dynamical structures 
and dark matter fractions of galaxies at $z\sim0.5{-}2$. 
Additionally, future observations of molecular gas with NOEMA and ALMA, and of ionized gas with VLT/ERIS and \JWST, 
will allow us to push these detailed dynamical studies to populations at $z\gtrsim3$.  
Finally, we will examine the non-circular motions of $z\sim1{-}2$ galaxies in detail, 
using existing and future high-resolution data. These high-resolution data will allow us to consider not only 
the general mass distributions in these galaxies, but to also constrain the internal dynamical processes that play key roles in 
the build-up and evolution of galaxies -- in particular, the build-up of bulges, which is expected to be rapid at this epoch, 
and that may be accompanied by the ``coring'' of the galaxies' dark matter halos. 

% ++++++++++++++++++++++++++++++++++++++++++++++++++++++++++++++
% ++++++++++++++++++++++++++++++++++++++++++++++++++++++++++++++

\acknowledgements

%%%%
We thank our colleagues at MPE, ESO-Garching and ESO-Paranal, LBT and IRAM, and members of the 
3D-HST, SINFONI/SINS \& zC-SINF and KMOS/KMOS$^{\mathrm{3D}}$ teams, 
who have contributed to, helped, or otherwise supported these observations and their analysis. 
We also thank the anonymous referee for their valuable feedback and suggestions that improved this manuscript. 
The data analyzed here are based on \Halpha observations collected with the integral field spectrometers SINFONI 
(SINS \& zC-SINF surveys, plus open time programs) 
and KMOS (KMOS$^{\mathrm{3D}}$ guaranteed time survey) 
obtained at the Very Large Telescope (VLT) of the European Southern Observatory (ESO), 
Paranal, Chile (under ESO programmes 073.B-9018, 074.A-9011, 075.A- 0466, 076.A-0527, 077.A-0527, 078.A-0600, 079.A-0341, 080.A-0330, 080.A-0339, 080.A-0635, 081.A-0672, 081.B-0568, 082.A-0396, 183.A-0781, 087.A-0081, 088.A-0202, 088.A-0209, 090.A-0516, 091.A-0126, 092.A- 0082, 092.A-0091, 093.A-0079, 093.A-0110, 093.A-0233, 094.A-0217, 094.A-0568, 095.A-0047, 096.A-0025, 097.A-0028, 098.A-0045, 099.A-0013, 0100.A-0039, 0100.A-0361, and 0102.B-0087). 
The analysis also includes CO observations within the PHIBSS1 and PHIBSS2 open time large projects, 
and the NOEMA$\mathrm{^{3D}}$ guaranteed time project, at the Northern Extended Array for Millimeter Astronomy 
(NOEMA, located on the Plateau de Bure) Interferometer of the Institute for Radio Astronomy in the Millimeter Range (IRAM),  
Grenoble, France. IRAM is supported by INSU/CNRS (France), MPG (Germany), and IGN (Spain). 
Finally, this work includes \Halpha slit spectroscopy obtained with the LUCI spectrometer at 
the Large Binocular Telescope (LBT) on Mount Graham, Arizona, USA. 
The LBT is an international collaboration among institutions in the United States, Italy, and Germany. 
LBT Corporation partners are: LBT Beteiligungsgesellschaft, Germany, representing the Max-Planck Society, 
The Leibniz Institute for Astrophysics Potsdam, and Heidelberg University; The University of Arizona on behalf 
of the Arizona Board of Regents; Istituto Nazionale di Astrofisica, Italy; The Ohio State University, 
and The Research Corporation, on behalf of The University of Notre Dame, University of Minnesota and University of Virginia. 
This work was supported in part by the Deutsche Forschungsgemeinschaft (DFG, German Research Foundation) 
within the Deutsch-Israelische Projektkooperation (DIP, German-Israeli Project Cooperation) 
under DFG/DIP grant STE/1869 2-1 / GE 625/17-1. 
TN acknowledges support by the Excellence Cluster ORIGINS which is funded by the DFG 
under Germany's Excellence Strategy -- EXC-2094 -- 390783311.

\software{Astropy\footnote{http://www.astropy.org}  \citep{astropy:2013, astropy:2018}, 
emcee \citep{Foreman-Mackey13}, 
corner \citep{corner}, 
IPython \citep{Perez07}, 
Matplotlib \citep{Hunter07}, 
MPFIT \citep{Markwardt09}, 
Numpy \citep{van2011numpy, 2020NumPy}, 
Pandas \citep{McKinney10, pandas2020}, 
Photutils \citep{photutils072, photutils101}, 
Scipy \citep{2020SciPy} } 

\facilities{VLT:Antu (KMOS), VLT:Yepun (SINFONI), IRAM:NOEMA, LBT (LUCI)}

% ++++++++++++++++++++++++++++++++++++++++++++++++++++++++++++++
% ++++++++++++++++++++++++++++++++++++++++++++++++++++++++++++++

\appendix

\vglue -20pt

%% ++++++++++++++++++++++++++++++++++++++++++++++++++++++++++++++
% ++++++++++++++++++++++++++++
\begin{figure*}[ht!]
\vglue -2pt
\centering
\includegraphics[width=0.94\textwidth]{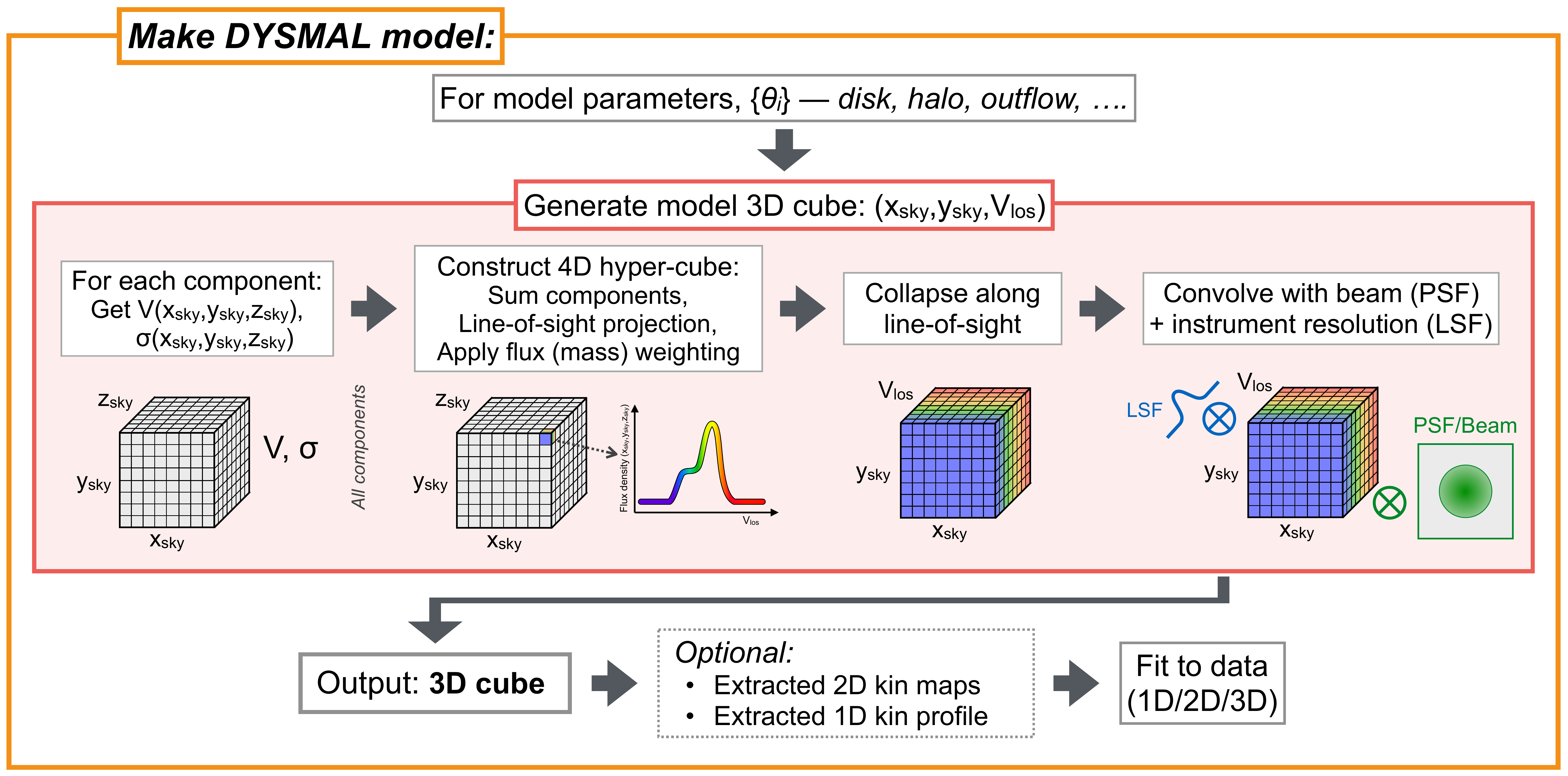}
\caption{
Schematic of how \Dysmalpy models are generated. 
The process involves a full forward modeling of the components in the sky frame $(x,y,z)$, 
construction of a composite 4D hyper-cube including line-of-sight projection and flux weighting, 
a collapse along the line-of-sight to determine $I_{\mathrm{intr.}}(x,y,V)$, 
and convolution with the instrument beam and response (i.e., PSF and LSF) to yield a 3D cube. 
This 3D model cube can then be directly compared to data cubes, or can be further processed 
for comparison to 1D kinematic profiles or 2D kinematic maps. 
}
\vspace{4pt} 
\label{fig:DYSMALschematic}
\end{figure*}
% ++++++++++++++++++++++++++++
%% ++++++++++++++++++++++++++++++++++++++++++++++++++++++++++++++

% ++++++++++++++++++++++++++++++++++++++++++++++++++++++++++++++
% ++++++++++++++++++++++++++++++++++++++++++++++++++++++++++++++
\section{\Dysmalpy: Dynamical Disk Modeling with DYSMAL}
\label{sec:appendixA}

DYSMAL is a code that uses a set of physical model mass and kinematic components to describe
and fit the kinematics of galaxies. 
Here, we discuss \Dysmalpy, which implements and extends the DYSMAL fitting models
that were introduced in \citet{Cresci09} and \citet{Davies11}, 
and includes subsequent improvements described in \citet{Wuyts16}, \citet{Genzel17}, and \citet{Ubler18}. 
Specifically, this new implementation in \texttt{python} now includes multiple halo models, 
outflow components, the ability to tie model component parameters together, 
and the choice of fitting using either least-squares minimization or Markov Chain Monte Carlo (MCMC) posterior sampling.

In this Appendix, we describe how a DYSMAL model is used to forward-model a full 3D mock cube 
$I_{\mathrm{mod}}(x_{\mathrm{sky}},y_{\mathrm{sky}},\Vlos)$  
that captures the composite kinematics, and accounts for all observational effects 
(including beam smearing and instrumental line broadening). 
This model cube can be either retained for 3D cube comparisons, 
or can be directly compared to 2D or 1D kinematic observations by 
extracting 2D maps or 1D profiles following the same procedure that was applied to the observed data. 
We then describe the procedures for fitting the DYSMAL model to observed data, either using MCMC or least-squares fitting.

As the DYSMAL model creation fully forward-models the galaxy kinematics, 
it is possible to directly fit for the intrinsic galaxy properties. 
Furthermore, because DYSMAL relies directly on mass distribution and kinematic profile parameterizations, 
these models allow for direct exploration of mass decomposition and dark matter fractions in galaxies, 
as well as any degeneracies or uncertainties in these physical quantities, 
as opposed to non-parametric kinematic fitting that 
requires further steps to interpret the recovered intrinsic galaxy kinematics.\footnote{Other 
approaches to kinematic fitting that use forward modeling to directly account for observational effects 
include GalPaK$^{\mathrm{3D}}$ (another parametric approach; \citealt{Bouche15}) 
and $^{\mathrm{3D}}$BAROLO (non-parametric, tilted ring modeling; \citealt{DiTeodoro15}).}

% ++++++++++++++++++++++++++++++++++++++++++++++++++++++++++++++
\subsection{Definition of DYSMAL Kinematic Models}
\label{sec:DYSMALmodels}

At their core, DYSMAL models are a composite set of mass and kinematic components, 
together with other galaxy properties, which together describe the mass profile, light profile, 
other kinematic components, and geometry of a galaxy. 
These components, along with galaxy and instrument parameterizations, 
are used to compute a full composite mock 3D cube $I_{\mathrm{mod}}(x_{\mathrm{sky}},y_{\mathrm{sky}},\Vlos)$  
that is comparable to real observed integral field spectrograph cubes.
The procedure for producing a mock 3D cube from the galaxy components and parameterizations is as follows. 
We also outline the major steps in the process in Figure~\ref{fig:DYSMALschematic}, 
and in Table~\ref{tab:dysmalpy_components} present an overview of the possible model components that can be used with \Dysmalpy.

We begin by defining an intrinsic coordinate system $(\xgal, \ygal, \zgal)$ for the galactic system. 
We define \xgal, \ygal as the position within the galaxy midplane 
(so $\Rgal=\sqrt{\xgal^2 + \ygal^2}$ is the radial distance from the rotational axis), 
and \zgal to be the vertical position (which is parallel to the rotational axis).
The \ygal axis is taken to be the axis about which the galaxy is inclined on the sky 
(so the projected major axis coincides with the \ygal axis).

The galactic system is then inclined at an angle $i$ relative to the line of sight (with $i=0^{\circ}, 90^{\circ}$ corresponding 
to face-on and edge-on orientations, respectively), and 
the major axis (blue side; \ygalhat) is oriented at an angle $\PA$ counter-clockwise from the upward direction in the 
observations (e.g., often angle East of North). 
The transformation from the sky coordinates 
back to the intrinsic galaxy coordinates is then

\vspace{-10pt}
\begin{align}
\xgal &= \left[(\xsky-\xcenter)\cos\PA - (\ysky-\ycenter)\sin\PA \right]\cos i \nonumber \\ 
& \qquad\ - \zsky \sin i \nonumber \\
\ygal &= \ \, (\xsky-\xcenter)\sin\PA + (\ysky-\ycenter)\cos\PA \nonumber \\
\zgal &=  \left[(\xsky-\xcenter)\cos\PA - (\ysky-\ycenter)\sin\PA \right]\sin i \nonumber \\ 
& \qquad\  + \zsky \cos i , 
\label{eq:galcoord}
\end{align}
where $(\xcenter,\ycenter)$ is the center of the galaxy within the observation field of view.

Next, for each mass component in the model, the mass distribution 
is used to determine the circular velocity within the midplane, 
$v_{\mathrm{circ,comp}}(\Rgal, \zgal=0)$. 
If the component is flattened, the modified circular velocity curve is included here. 
The composite circular velocity curve is then found by summing in quadrature: 

\vspace{-5pt}
\begin{equation}
\vcirctot^2(\Rgal) = \sum_{\mathrm{comp}\ i} v_{\mathrm{circ,}i}^2(\Rgal).
\end{equation}
Alternatively, if adiabatic contraction of the halo component is to be included, the total circular velocity is instead 
calculated using the implicit equation from \citet{Burkert10} (Eqs.~18 \& 19):

\vspace{-10pt}
\begin{align}
&\vcirctot^2(\Rgal) = v_{\mathrm{bary,tot}}^2(\Rgal) + v_{\mathrm{halo}}^2(R'),  \nonumber \\ 
&R' = \Rgal \left[ 1 + \frac{\Rgal \times v_{\mathrm{bary,tot}}^2(\Rgal)}{R' \times v_{\mathrm{halo}}^2(R')} \right], 
\label{eq:adiabatic_contraction}
\end{align}
\vglue -15pt
\begin{align}
\mathrm{with} \quad 
&v_{\mathrm{bary,tot}}^2(\Rgal) = 
\raisebox{0.25\depth}{$\sum\limits_{\small\substack{\mathrm{baryonic}\\\mathrm{comp}\ i}}$} 
\left[  v_{\mathrm{circ,}i}^2(\Rgal)  \right] . \hspace{48pt} \nonumber
\end{align} 
For simplicity, we assume this total $\vcirctot(\Rgal)$ is independent of \zgal, 
or $\vcirctot(\Rgal,\zgal)=\vcirctot(\Rgal,0)$ (i.e. cylindrical shells of constant circular velocity).

%%%%%%%%%%%%%%%%%%%%%
%%%%%%%%%%%%%%%%%%%%%
\begin{table}
\scriptsize
\tablenum{5} 
 \centering \caption{Possible Model Components in \Dysmalpy}
 \setlength{\tabcolsep}{0.06in} 
\label{tab:dysmalpy_components}
\begin{tabular*}{0.9\columnwidth}{l @{\extracolsep{\fill}} c}
\hline
\hline  
& \\[-8pt]
\textbf{\underline{Components}} & \textbf{\underline{Key Parameters}}  \\
Black hole & $M_{\mathrm{BH}}$ \\
Freeman disk & $\Sigma_0, r_d$ \\[2pt]
S\'ersic (flattened or spherical) & $n_S, q_0, \re$ \\
\hskip 15pt \textit{e.g.: Disk + bulge} & $\nSdisk, \qint, \redisk$; \\
& $\nSbulge, q_{0,\mathrm{bulge}}, \rebulge$ \\ 
Dark matter halo & \\
\hskip 15pt \textit{NFW} & \lMvir or \fDM, \chalo \\
\hskip 15pt \textit{Two-power halo} & $\alphainner, \beta, \lMvir, \chalo$ \\  
\hskip 15pt \textit{Burkert} & \rB, \lMvir \\
\hskip 15pt \textit{Einasto} & \alphaEinasto, \lMvir, \chalo \\ 
\hskip 15pt \textit{Dekel-Zhao} & \lMvir, $s_1$, $c_2$ \\[2pt]
Intrinsic velocity dispersion & \\
\hskip 15pt \textit{Isotropic} & \sigmaint \\[2pt]
Biconical Outflow &  $\theta_{\rm inner}$, $\theta_{\rm outer}$, $i_\mathrm{out}$, $r_\mathrm{end}$, $n$, $\tau$, $A$,  
$v_\mathrm{max}$ \\[2pt]
Radial motion / flow & \\
\hskip 15pt \textit{Constant radial flow} & $v_{r}$\\[5pt]
\hline
\end{tabular*}
\end{table}
%%%%%%%%%%%%%%%%%%%%%
%%%%%%%%%%%%%%%%%%%%%

Mass components that are implemented in DYSMAL include S\'ersic components 
with or without flattening (following \citealt{Noordermeer08}), 
which can be used to describe disk (e.g., flattened and \hbox{$n_S\sim1$}) and bulge 
(e.g., spherical or flattened and \hbox{$n_S\sim4$}) baryonic components. 
Tables with pre-computed rotation curves following \citet{Noordermeer08} 
have been calculated for $n_S=0.5,\ldots,8$ in steps of $0.1$, with inverse $1/q_0=[1,2,3,4,5,6,8,10,20,100]$. 
The pre-computed rotation curves with the nearest $n_S$ and $q_0$ are then 
used when including a deprojected (flattened or spherical) S\'ersic mass component 
(i.e., when $\nSdisk < 0.5$, $n_S=0.5$ rotation curves are used, but $\nSdisk$ is used for the light distribution; see below). 
Also included are halos following NFW (\citealt{Navarro96}), Two-Power (\citealt{BT08}), Burkert (\citealt{Burkert95}), 
Einasto (\citealt{Einasto65}), and Dekel-Zhao (\citealt{Dekel17}, \citealt{Freundlich20}) profiles.

The intrinsic dispersion profile of the model, $\sigma(\xgal,\ygal,\zgal)$, can be parameterized in different ways. 
The simplest option is a constant, isotropic intrinsic dispersion, $\sigma(\xgal,\ygal,\zgal) = \sigmaint$, 
but other possible options include parameterizations $\sigma(\Rgal)$ 
based on the surface density of the galactic disk.

For thick disks, as commonly seen in high-redshift galaxies, 
part of the total dynamical support is from pressure support, and not just from rotation. 
Thus, the rotation velocity of such a galaxy is lower than the circular velocity, 
as a result of asymmetric drift. 
We describe the rotation velocity of our system, with asymmetric drift included, 
following the formulation of \citet{Burkert10}, 
derived for a self-gravitating exponential disk and constant dispersion $\sigmaint$: 

\vspace{-10pt}
\begin{align}
\vrot^2(\Rgal)  &= \vcirctot^2(\Rgal) - 2 \, \sigmaint^2   \left(\frac{\Rgal}{r_d}\right) 
\label{eq:asymm_drift}
\end{align} 
where $r_d$ is the disk scale radius, and $r_d = \re/1.68$ for an exponential disk. 
If we follow the derivation by \citet{Burkert10}, but assume the 
self-gravitating disk follows a more general 
S\'ersic distribution, 
the asymmetric drift correction is then

\vspace{-15pt}
\begin{align}
\vrot^2(\Rgal) = 
 \vcirctot^2(\Rgal) - 2  \, \sigmaint^2  \left(\frac{b_{n_S}}{n_S}\right) \left(\frac{\Rgal}{\re}\right)^{1/n_S}, 
\label{eq:asymm_drift_sersic}
\end{align} 
where $2\,(b_{n_S}/n_S)=3.36$ for an exponential disk ($n_S=1$). 
Even more generally, assuming the dispersion is constant with radius \Rgal (but not necessarily with height $z$ above the midplane), 
then the asymmetric drift correction is given by Eq. 3 of \citet{Burkert10} (with a correction term of $\sigmaint^2(d\ln\rho/d\ln{}r)$, 
where $r=\Rgal$).

A radial motion component can be imprinted on top of the galaxy rotational motion, 
$\vec{v}_{\mathrm{radial\ motion}}(\xgal, \ygal, \zgal)$. 
One option is to assume a constant radial flow, defined as 
$\vec{v}_{\mathrm{radial\ motion}}(\xgal, \ygal, \zgal) = -v_{r} \, \hat{r}_{\mathrm{gal}}$ 
(where $r_{\mathrm{gal}}^2 = \Rgal^2 + \zgal^2$ is the 3D radius of the galaxy, and positive $v_r$ 
corresponds to inflows), 
but profiles that vary with radius or azimuthal angle could also be implemented in the future.  
In the simplest case, we assume this is a perturbation on the other motions, so this 
motion component follows the light distribution and geometry of the galaxy.

In addition, a detailed biconical outflow model similar to the one described in \citet{Bae:2016aa} is implemented. 
For this model component, the light and kinematics follow two axisymmetric cones 
that share an apex at the location where the outflow is launched, $(x_{\rm out,0}, y_{\rm out,0})$. 
The shape of the cones are primarily defined by two opening angles, $\theta_{\rm inner}$ and $\theta_{\rm outer}$, 
which delineate the walls of the cones. Each opening angle is measured 
from the outflow axis such that $0^{\circ}$ is along the outflow axis. 
Only regions between $\theta_{\rm inner}$ and $\theta_{\rm outer}$ then 
produce line emission and affect the line-of-sight (LOS) kinematics. 
Finally, the cones have a maximum radial extent out to $r_\mathrm{end}$.

Relative to the plane of the sky, the cones have an inclination, $i_{\rm out}$, and position angle $\PA_{\rm out}$ 
that can be different from the galaxy. $i_{\rm out}$ is defined such that $0^{\circ}$ indicates an outflow axis along the LOS. 
The transformation from the sky coordinates $(\xsky,\ysky,\zsky)$ back to the intrinsic outflow coordinates $(\xoutf,\youtf,\zoutf)$ 
is then exactly the same as in Equation~\ref{eq:galcoord} except with the specific outflow central coordinate, inclination, and $\PA$.

Three choices of outflow velocity radial profiles are possible: `increasing', `decreasing', or `both'. 
Each of these is parameterized in the following way:
\begin{align}
    \mathrm{Increasing:}\, &v_{\rm out}(r) = v_{\rm max}\left(\frac{r}{r_{\rm end}}\right)^n \nonumber \\
    \mathrm{Decreasing:}\, &v_{\rm out}(r) = v_{\rm max}\left[1 - \left(\frac{r}{r_{\rm end}}\right)^n\right] \nonumber \\
    \mathrm{Both:}\, &v_{\rm out}(r) = 
        \begin{cases}
        v_{\rm max}\left(\frac{r}{r_{\rm turn}}\right)^n,& r < r_{\rm turn} \\
        v_{\rm max}\left(2 - \frac{r}{r_{\rm turn}}\right)^n,& r > r_{\rm turn}
        \end{cases}
\end{align}
\noindent where $n$ is the power law index of the radial profile, $v_{\mathrm{max}}$ defines the maximum velocity that occurs at 
$r = r_\mathrm{end}$ for `increasing', $ r = 0$ for `decreasing', and $r = r_\mathrm{turn}$ for `both'. 
A velocity dispersion profile $\sigma_{\mathrm{out},0}(r)$ can be chosen for 
the outflowing gas in the same way as for the galaxy mass components.

Next, the LOS velocity and dispersion cubes are constructed, and sampled over the sky coordinate frame. 
For the composite rotation velocity profile, the LOS projection factor is  $\cos\psi \sin i =  \sin i \; (\ygal/\Rgal)$. 
Similarly, LOS projection is determined for non-isotropic dispersion profiles and 
outflow or radial motion/inflow components.\footnote{I.e., as $\left[\vec{\sigma}(\xgal,\ygal,\zgal)\cdot\hat{z}\right]$ or 
$\left[\vec{v}_{\mathrm{radial\ motion}}(\xgal,\ygal,\zgal)\cdot\hat{z}\right]$.} 
The LOS velocity and dispersion cubes are thus:

\vspace{-15pt}
\begin{align}
\! 
V(\xsky, &\ysky,\zsky) = \vrot(\Rgal)  \left(\frac{\ygal}{\Rgal}\right) \sin i  \nonumber \\
& + \left[v_{\mathrm{radial\ motion}}(\xgal,\ygal,\zgal)\right]_{\mathrm{LOS}}  + \Vsys  
\label{eq:vel_cube}
\end{align}
\vglue -15pt
\begin{align}
\sigma(\xsky,\ysky,\zsky) &= \left[\sigma(\Rgal)\right]_{\mathrm{LOS}} , \hspace{59pt} 
\label{eq:disp_cube}
\end{align}
\noindent 
where Eq.~\ref{eq:galcoord} is used to convert from sky to galaxy coordinates, 
and \Vsys is the systemic velocity of the system.

For the biconical outflow component, these cubes are: 
\begin{align}
V_{\mathrm{out}}(\xsky,\ysky,\zsky)  &=  v_{\mathrm{out}}(r_\mathrm{out})\cos i^{\prime} + \Vsys 
\label{eq:velout_cube}
\end{align}
\vglue -15pt
\begin{align}
\sigma_\mathrm{out}(\xsky,\ysky,\zsky) &= \left[\sigma_{\mathrm{out},0}(r_\mathrm{out})\right]_{\mathrm{LOS}} , \hspace{12pt} 
\label{eq:dispout_cube}
\end{align} 
where $r_\mathrm{out} = \sqrt{\xoutf^2 + \youtf^2 + \zoutf^2}$, and $i^{\prime}$ in Eq.~\ref{eq:velout_cube} 
ranges from $i_\mathrm{out} - \theta_\mathrm{outer}$ to $i_\mathrm{out} + \theta_\mathrm{outer}$ 
depending on the location within the outflow cone.

In order to construct a full model cube $I(\xsky,\ysky,\Vlos)$, 
the information from the velocity and dispersion cubes $V(\xsky,\ysky,\zsky)$ and $\sigma(\xsky,\ysky,\zsky)$ 
must be combined and collapsed along the line of sight (\zsky). 
The composite velocity profile at a fixed position (\xsky,\ysky) is the intensity-weighted sum of 
all kinematic components along the line of sight. 
This requires parameterization of the light distribution of the galactic system, $f(\xgal,\ygal,\zgal)$, 
transformed to sky coordinates $(\xsky,\ysky,\zsky)$ using Eq.~\ref{eq:galcoord}.

In practice, one option is to assume one (or more) mass component (e.g., the baryonic disk) emits light, 
with a constant mass-to-light ratio $\Upsilon$. 
This can be approximated by assuming a 2D S\'ersic flux distribution 
within the galaxy midplane, combined with a Gaussian profile in the \zgal direction with a width related to the 
assumed S\'ersic flattening $q_0$ and the component \re. 
In this case, $f(\Rgal,\zgal) = \Sigma_0 \exp[-b_{n_S}(\Rgal/\re)^{1/{n_S}}]\, \times \, \exp[-0.5(\zgal/h_z)^2]$,
with $h_z = q_0 \re / 1.177$.

The biconical outflow light distribution exponentially decreases as a function of radius with the following profile and parameterization:
\vspace{-5pt}
\begin{equation}
    f_\mathrm{out}(r_\mathrm{out}) = A\exp\left(-\frac{\tau r_{\rm out}}{r_{\rm end}}\right)
\end{equation}
\noindent where $A$ is the flux at $r_{\mathrm{out}}=0$, and $\tau$ controls the rate at which the flux declines.

The intensity-weighted kinematic distribution collapsed along the line of sight is determined 
by combining the LOS velocity, dispersion, and flux cubes, assuming the profile at each position 
can be described as a Gaussian in velocity $\Vlos$ of total flux $f(\xsky,\ysky,\zsky)$ 
and dispersion $\sigma(\xsky,\ysky,\zsky)$, centered at $V(\xsky,\ysky,\zsky)$:

\vspace{-5pt}
\begin{align}
I&_{\mathrm{intr.}}(\xsky,\ysky,\Vlos) = \nonumber \\
&\sum_{\zsky} \, \Biggggl( \, \frac{f(\xsky,\ysky,\zsky)}{\sigma(\xsky,\ysky,\zsky) \sqrt{2\pi}} \ \times  \nonumber\\
&\qquad \qquad  \exp \, \Bigggl\{-\frac{\left[\Vlos-V(\xsky,\ysky,\zsky)\right]^2}{2\sigma(\xsky,\ysky,\zsky)^2}\Bigggr\} \ \ + \nonumber\\
&\qquad \ \frac{f_\mathrm{out}(\xsky,\ysky,\zsky)}{\sigma_\mathrm{out}(\xsky,\ysky,\zsky) \sqrt{2\pi}}  \  \times \nonumber\\
& \qquad \qquad   \exp \, 
\Bigggl\{-\frac{\left[\Vlos-V_\mathrm{out}(\xsky,\ysky,\zsky)\right]^2}{2\sigma_\mathrm{out}(\xsky,\ysky,\zsky)^2}\Bigggr\} \Biggggr),  
\end{align} 
\noindent
using the respective light, velocity, and velocity dispersion distributions for the outflow component and the galaxy.\footnote{Note that 
if there is an inflow signature from a separate gas component, this can be added to the model 
with a separate light distribution and geometry (i.e., not in Eq.~\ref{eq:vel_cube}),  
similar to the treatment of the outflow component.}

Finally, this intrinsic kinematic model cube is convolved with a spatial point spread function (PSF) 
and a spectral line spread function (LSF) to directly include the impact of instrumental and observational 
effects on the kinematic model. Specifically, the convolution is performed
with a 3D kernel folding together the spatial PSF and 
the spectral LSF: 

\vspace{-10pt}
\begin{align}
I_{\mathrm{mod}}(\xsky,\ysky,\Vlos) &=   I_{\mathrm{intr.}}(\xsky,\ysky,\Vlos)  \nonumber\\
&\quad \otimes \ \left[\mathrm{PSF}(\xsky,\ysky) * \mathrm{LSF}(\Vlos) \right] 
\label{eq:Imod}
\end{align}

This model cube is the end result of the DYSMAL forward modeling process.  
The cube can now be used for comparisons and fitting to observed kinematic data, 
which we discuss in Section~\ref{sec:DYSMALfitting}.

% ++++++++++++++++++++++++++++++++++++++++++++++++++++++++++++++
\subsection{Fitting Galaxy Kinematics using DYSMAL Models}
\label{sec:DYSMALfitting}

The DYSMAL models presented in the previous subsection can be used to fit a wide variety of 
galaxy mass and kinematic components to a range of observational data.

First, a DYSMAL model set should be selected based on the specific application. 
For instance, this might be comprised of a disk, bulge, and halo mass components together with an 
intrinsic velocity dispersion profile, or it may include a galaxy disk together with an outflow component. 
The parameters of each of these model components 
are then set to a fixed value (e.g., fixed bulge-to-total ratio, \bt), 
allowed to vary as free fit parameters (e.g., free total mass), 
or are determined as a function of some other parameter 
(e.g., fitting directly for $\fDM$, and then using this to find the halo mass for a NFW profile; 
alternatively, setting the halo mass from the total baryonic mass, using 
a fixed $\fgas$ and a particular stellar mass-halo mass relation).

Next, depending on the type of observational data, extractions may need to be made from the DYSMAL model cubes. 
If the observations are to be fit in 3D, using IFS cubes, then the 3D mock DYSMAL cubes 
are directly comparable to the data. 
Alternatively, the models can be used to fit 2D extracted kinematic maps or 
1D extracted kinematic profiles (or slit observations), 
for instance fitting both velocity $V$ and dispersion $\sigma$ simultaneously 
(though other possibilities are to fit only $V$, or to fit all of $V$, $\sigma$, and the flux distribution). 
This is accomplished by applying the same extraction methodology used for the observations 
on the mock cubes (e.g., Gaussian-fit extractions to 2D velocity and dispersion maps; or 
flared or straight slit aperture extraction to 1D profiles).

We stress that all observational effects (including beam-smearing, the instrumental line spread function, 
and any extraction from 3D) are directly included in the resulting 1D/2D/3D DYSMAL models. 
The full forward modeling of this procedure therefore allows us to directly fit for the intrinsic model properties.

In \Dysmalpy, fitting can be performed either with 
Markov Chain Monte Carlo (MCMC) parameter space exploration (using \texttt{emcee}; \citealt{Foreman-Mackey13}) 
or least-squares fitting (using \texttt{MPFIT}; \citealt{Markwardt09}). 
For MCMC fitting with \texttt{emcee}, priors $\log p(\theta_i)$ are first selected 
for each free parameter $\theta_i$ in the DYSMAL model; 
options include flat or Gaussian priors (either bounded or unbounded). 
The likelihood function for our model is defined to be a Gaussian distribution. 
For 3D fitting, 
\begin{align}
\log \mathcal{L} = &-0.5  \left(\frac{1}{f}\right)
\sum_{\xsky,\ysky,\Vlos}  m(\xsky,\ysky,\Vlos) \ \times    \nonumber \\
& \Bigggl\{ \ 
 \Bigggl[ \frac{ I_{\mathrm{obs}}(\xsky,\ysky,\Vlos)
-I_{\mathrm{mod}}(\xsky,\ysky,\Vlos)} {\mathrm{err}_{{\mathrm{obs}}(x,y,V)}  \left[w(\xsky,\ysky,\Vlos)\right]^{-1/2} }  \Bigggr]^{\,2}
 \nonumber \\ 
& \qquad + 
\log \ \Bigggl[\frac{2\pi \big(\mathrm{err}_{{\mathrm{obs}}(\xsky,\ysky,\Vlos)}\big)^2}{w(\xsky,\ysky,\Vlos)}\Bigggr] \ \Bigggr\}\, , 
\end{align} 
where $I_{\mathrm{mod}}$ is from Eq.~\ref{eq:Imod}, 
$I_{\mathrm{obs}}$ is the observed 3D cube, 
$m(\xsky,\ysky,\Vlos)$ is a data mask (e.g., removing bad pixels or low S/N regions), 
$w(\xsky,\ysky,\Vlos)$ is any weighting to be applied to the cube (e.g., effectively modifying the uncertainties), 
and $f$ is an optional factor to account for oversampling of the data relative to the spatial and spectral resolution of the 
data (to approximate the number of independent data points, 
removing any impact of oversampling on the relative importance of the likelihood and priors). 
For 2D or 1D fitting, 
\begin{align}
&\log \mathcal{L} = -0.5   \left(\frac{1}{f}\right) \times \sum_{\substack{X=V,  
[\sigma],  \\ \quad [\mathrm{flux}]}}
  \nonumber \\
& \Bigggl\{ \sum_{i} m_i \left[ w_i \left( \frac{X_{\mathrm{obs},i} - X_{\mathrm{mod},i}}{\mathrm{err}_{X,i}} \right)^2 
+ \log\left(\frac{2\pi \ \mathrm{err}_{X,i}^2}{w_i} \right)
\right] \Bigggr\}, 
\label{eq:loglike1D2D}
\end{align} 
where $X$ is either the velocity $V$, dispersion $\sigma$, or flux
(e.g., 1D profiles along the slit $V(p)$, $\sigma(p)$, $I(p)$, 
or 2D maps of $V$, $\sigma$ and line intensity $I$), and 
$X_{\mathrm{obs}}$ and $X_{\mathrm{mod}}$ 
are the data and extracted model maps/profiles, respectively. 
The outer sum over the maps/profiles $X$ includes only the maps/profiles that are being used for fitting 
(e.g., simultaneously fitting velocity $V$ and dispersion $\sigma$; see discussion above), 
and the sum over $i$ denotes the sum over all pixels or data points in the maps/profiles. 
Similar to the 3D log likelihood, $m_i$ is the data mask, $w_i$ is any data weighting (or no weighting, with $w_i=1$), 
and $f$ is the optional factor accounting for spatial data oversampling.  
The log posterior probability is then 

\vspace{-10pt}
\begin{equation}
\log P(\theta | \mathrm{obs}) = \log \mathcal{L}( \mathrm{obs} | \theta )  + \log p(\theta) + \mathrm{const}, 
\label{eq:logprob}
\end{equation}
with the log likelihood as defined above and taking
the prior as the composite of the individual parameter priors,  $\log p(\theta) = \sum_i  \log p(\theta_i)$.

The ``best-fit'' parameter values are determined as the maximum a posteriori (MAP) values from the posteriors, 
with the option to jointly analyze the posteriors of parameters which exhibit degeneracies, to 
ensure the MAP values correspond to high probability regions of the multidimensional posterior distribution. 
The upper and lower $1\sigma$ uncertainties for each parameter are estimated independently, 
using the shortest 68\% interval of the marginalized posterior (to ensure reasonable uncertainty estimates in cases 
where the marginalized posterior is peaked near a boundary for the parameter).

Alternatively, fitting can be performed with \texttt{MPFIT}, which 
uses the Levenberg-Marquardt technique to perform least-squares fitting. 
For the least-squares fitting with \Dysmalpy, we define 
$\chi^2_{\mathrm{kin}}$ that is tailored to the data dimensionality. 
For 3D fitting, 

\vspace{-10pt}
\begin{align}
\chi^2_{\mathrm{kin}} = 
&\sum_{\xsky,\ysky,\Vlos}  m(\xsky,\ysky,\Vlos)  \; w(\xsky,\ysky,\Vlos)  \nonumber \\
&\times\left[ \frac{I_{\mathrm{obs}}(\xsky,\ysky,\Vlos) - 
I_{\mathrm{mod}}(\xsky,\ysky,\Vlos)}{\mathrm{err}_{I_{\mathrm{obs}}(\xsky,\ysky,\Vlos)}} \right]^2
\end{align} 
where as before, $I_{\mathrm{obs}}$ and $I_{\mathrm{mod}}$ are the observed and model 3D cubes, respectively, 
$m(\xsky,\ysky,\Vlos)$ is a mask based on the data (e.g., removing bad pixels or low S/N regions) and
$w(\xsky,\ysky,\Vlos)$ is any weighting to be applied to the cube. 
For 2D or 1D fitting, 

\vspace{-10pt}
\begin{equation}
\chi^2_{\mathrm{kin}} =            
   \sum_{\substack{X=V, [\sigma],  
   \\ \quad [\mathrm{flux}]}} 
\left[ \sum_{i} m_i \,  w_i  \left( \frac{X_{\mathrm{obs},i} - X_{\mathrm{mod},i}}{\mathrm{err}_{X,i}} \right)^2   \right], 
\end{equation} 
where again $X_{\mathrm{obs}}$ and $X_{\mathrm{mod}}$ are the observed and model 
velocity, dispersion or flux maps/profiles
(e.g., the observed $V(p)$, $\sigma(p)$, and flux $I(p)$ along the slit for 1D profiles), 
the outer sum only includes the observed maps/profiles used for simultaneous fitting 
(e.g., both $V$ and $\sigma$, or all of $V$, $\sigma$, and the flux distribution), 
and $m_i$ and $w_i$ are the data mask and any weighting of the data, respectively.

The best-fit parameter values are then taken directly from the least-squares minimization solution. 
One approach to derive uncertainties for the least-squares fitting 
is to then sample the value of $\chi^2_{\mathrm{kin}}$ over a grid of values for all free parameters, 
and use this to determine the uncertainty intervals. 
However, in practice it is often more computationally efficient to estimate the uncertainties through 
an accompanying MCMC exploration of the posterior distribution.

% ++++++++++++++++++++++++++++++++++++++++++++++++++++++++++++++
\subsection{Importance of Prior Choice in MCMC Fitting}
\label{sec:MCMCprior}

While MCMC fitting provides a number of benefits (e.g., simultaneously enabling uncertainty estimation), 
it is crucial to consider how the choice of priors impacts the sampled posterior distribution. 
In MCMC sampling, priors can help to restrict fits with degeneracies by 
applying constraints from ancillary information or physical feasibility (e.g., through bounding 
and/or the application of Gaussian priors), so that 
the posterior distribution reflects both the likelihood from the data and these prior constraints. 
Parameter priors can also be uniform, so the posterior is driven by the likelihood function. 
However, even adopting uniform priors is a choice that impacts later analysis, 
because the 
sampled posterior distribution also depends on how the fit was \textit{parameterized}.\footnote{For further 
discussion on the impact of parameter transformations on probability (i.e., prior and posterior) distributions, 
see \citet{Sivia06}, \citet{Hogg12}, and \citet{Hogg18}.}

The impact of priors is particularly important when analyzing parameters \textit{inferred} from the fit values. 
While not fit directly, there is an ``effective prior'' imprinted on the distribution of sample values for these derived parameters. 
This ``effective prior'' is determined by the choice of free fit parameters and their chosen priors (even uniform priors),  
and depends on the relationship between the fit and derived parameters. 
Formally, this is simply a change of variables in a probability distribution function (pdf), 
going from the fit parameter's prior pdf to a transformed pdf. 
If $x$ and $y$ are related through $y=f(x)$ and $x$ is fit with a uniform, bounded prior, 
then as the total probability must be conserved,\footnote{I.e., $\int_{a}^{b} p_X(x) \, dx = \int_{f(a)}^{f(b)} p_Y(y) \, dy$}
the effective prior on $y$ 
will be proportional to  $\abs{\frac{d}{dy} \left[ f^{-1}(y) \right]}$, 
where $f^{-1}$ is the inverse function (\citealt{Casella02}, Eq. 2.1.10).\footnote{More generally, 
for a prior pdf $p_X(x)$ for $x$, the transformed prior pdf for $y=f(x)$ is 
$p_{Y}(y) = p_{X}\left( f^{-1}(y) \right) \abs{ \frac{d}{dy} \left[ f^{-1}(y) \right] }$, 
assuming ``well-behaved'' functions (i.e., $p_X(x)$ is continuous and normalizable, 
and $f^{-1}(y)$ is continuously differentiable; \citealt{Casella02}, Sec 2.1, Theorem 2.1.5).}

The issue of free parameter and prior selection can have a large impact when 
fitting galaxy kinematics using MCMC sampling with \Dysmalpy. 
For example, for a fit with a baryonic component and a NFW halo, 
if \lMvir is chosen as the free parameter with a flat prior, 
then the distribution of $\fDM=v_{\mathrm{DM}}^2(\re)/v_{\mathrm{circ}}^2(\re)$ for the MCMC sampling 
(calculated as ``blobs'' from the \texttt{emcee} sampler) will often diverge towards 0 and 1. 
This arises in part because a flat prior on \lMvir is equivalent to a prior for \fDM that diverges 
as $\fDM \to 0$, or $\fDM \to 1$, because sampling uniformly from the \lMvir prior produces 
a pile-up of \lMvir values that all map to similarly small/large values of \fDM. 
If \fDM is fairly poorly constrained by the data (i.e., a shallow likelihood function), 
the effective probability distribution for \fDM will thus primarily reflect this ``effective,'' diverging prior.

%% ++++++++++++++++++++++++++++++++++++++++++++++++++++++++++++++
% ++++++++++++++++++++++++++++
\begin{figure*}[ht!]
\centering
\includegraphics[width=0.65\textwidth]{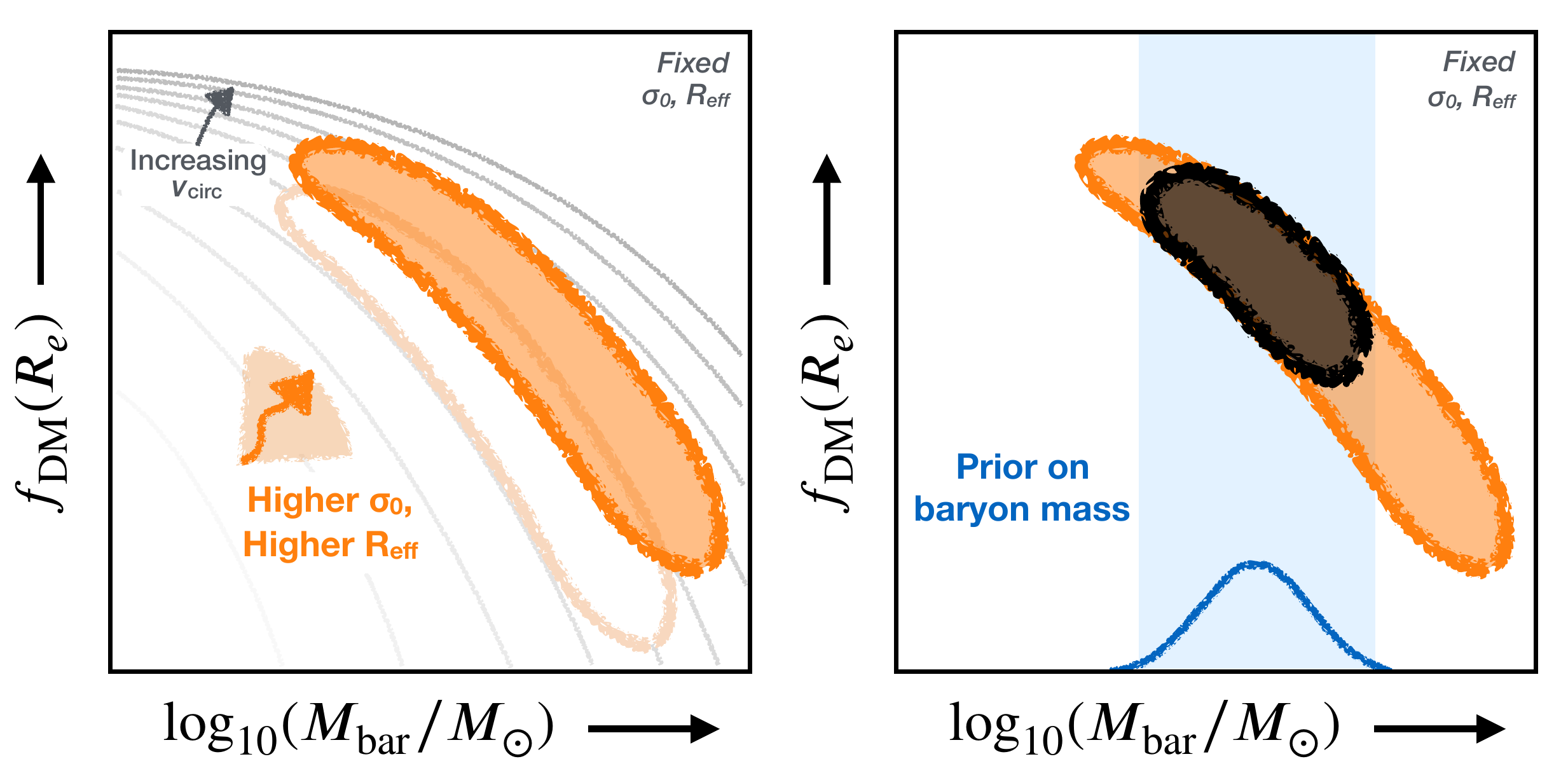}
\vglue -6pt
\caption{
Illustration of how the velocity dispersion and disk radius, and the prior on the baryonic mass, 
can impact the degeneracy between a galaxy's total baryonic mass and its dark matter fraction. 
There tends to be an anti-correlation between \fDM and \lMbar at constant \sigmaint and \redisk, 
roughly tracing lines of constant \vcirc (i.e., total dynamical mass). 
(Left panel) 
If the dispersion or disk radius is increased, then the likelihood degeneracy region shifts towards 
higher \fDM and higher \lMbar (as illustrated in the left panel). 
The exact direction and magnitude of the shifts depend on the detailed model values, 
but in general a higher \vcirc is needed to counterbalance the increased asymmetric 
drift correction or stretched rotation curve (for \sigmaint and \redisk, respectively). 
(Right panel) For a given fixed \sigmaint and \redisk, by applying a Gaussian prior on the baryonic mass, 
the posterior \fDM--\lMbar degeneracy (dark blue region, right panel) will be modified 
relative to the likelihood degeneracy (purple region). 
}
\vspace{6pt} 
\label{fig:fdm_mbar_degen_illustration}
\end{figure*}
% ++++++++++++++++++++++++++++
%% ++++++++++++++++++++++++++++++++++++++++++++++++++++++++++++++

Therefore, if the aim is to measure \fDM rather than \lMvir, it is better to fit directly for \fDM. 
For this inverted parameterization, the choice of a flat prior for \fDM corresponds to 
an effective prior on \lMvir that is a peaked distribution
(i.e., essentially the inverse of the opposite case that diverges towards the bounds). 
Because of this parameterization and prior choice issue, in this paper we chose
to fit directly for \fDM, as this is the quantity of interest.

%% ++++++++++++++++++++++++++++++++++++++++++++++++++++++++++++++
% ++++++++++++++++++++++++++++
\begin{figure*}[t!]
\centering
\includegraphics[width=0.75\textwidth]{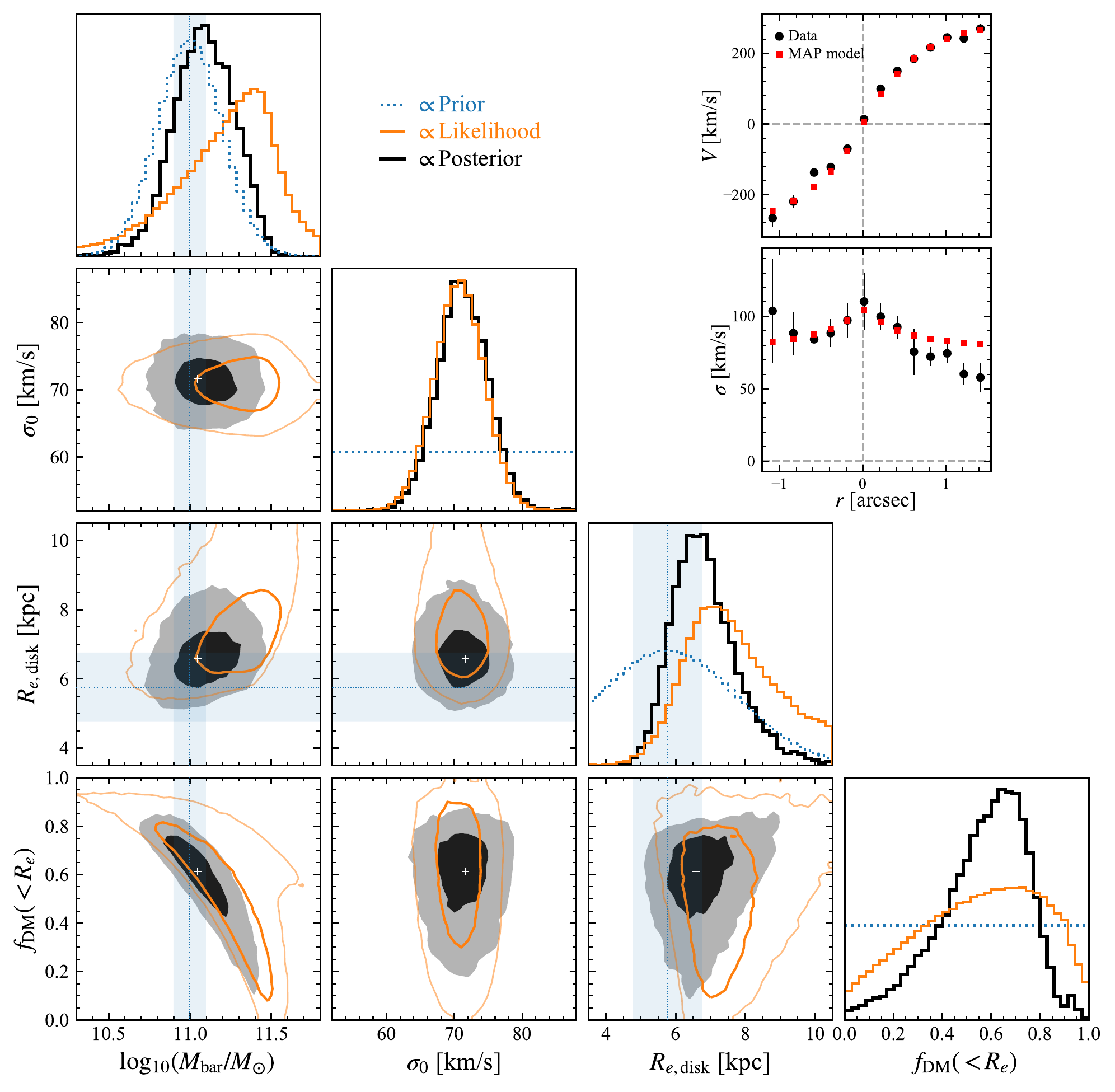}
\caption{
Comparison of the marginalized prior, posterior, and approximate likelihood contours 
for the free parameters in the 1D fitting for one galaxy in our sample (BX482). 
The normalized priors for each parameter are shown in the histogram panels (diagonal) as blue dotted curves. 
For Gaussian priors, the center and standard deviation are marked with vertical/horizontal blue lines and shaded regions, respectively. 
The marginalized approximate likelihood and posterior distributions are shown in orange and black/grey, respectively, 
with the 1 and 2 $\sigma$ intervals denoted with the 2D contours 
(with a 2D Gaussian filter of standard deviation 0.5 bins applied for clarity).  
The Gaussian priors on \lMbar and \redisk impact the peak and broadness of the posterior distribution 
for both parameters, and additionally contribute to a narrower \fDM--\lMbar degeneracy, both along the anti-correlation (\lMbar)
and perpendicular to it (\redisk). 
The \textit{maximum a posteriori} (MAP) values of each parameter (found by joint posterior analysis) are shown with white plus signs. 
We also show the 1D observed $V(r)$ and $\sigma(r)$ curves along with the 
1D-extracted MAP model (black circles, red squares, respectively) in the upper-right inset panels. 
}
\vspace{6pt} 
\label{fig:PLP_corner}
\end{figure*}
% ++++++++++++++++++++++++++++
%% ++++++++++++++++++++++++++++++++++++++++++++++++++++++++++++++

% ++++++++++++++++++++++++++++++++++++++++++++++++++++++++++++++
% ++++++++++++++++++++++++++++++++++++++++++++++++++++++++++++++
\subsection{DYSMAL Parameter Fitting Degeneracies for\\RC41 Curves and the Impact of Priors}
\label{sec:dysmal_degen_plp}

Fitting the kinematics of high-redshift galaxies with mass and kinematic models --- such as with DYSMAL --- 
is complicated by degeneracies between different components. 
These degeneracies arise in part from the impact of beam smearing, 
given the relatively low spatial resolution of the observations, 
and from relatively modest S/N of the data, 
even for very deep observations such as our RC41 sample.  
At lower S/N and lower spatial resolution, it can be difficult to disentangle the velocity signatures 
of a bulge, thick disk, and halo, 
as profile differences can be smoothed out to the point where they 
cannot be distinguished within the observational uncertainties.\footnote{We note 
that while we focus here on how these issues impact parametric kinematic modeling, 
these resolution and S/N limitations of the data also complicate constraints of galaxy kinematics using 
nonparametric methods.} 
Thus, while the lower halo concentrations at higher redshifts should help to break the disk-halo degeneracy 
that is observed in local galaxies (see the discussion in \citetalias{Genzel20}), 
the relatively limited spatial resolution and S/N of our observations compared to what is currently achievable for 
local galaxies tends to produce a similar baryon-halo degeneracy.

When modeling our galaxies with a bulge, a thick disk, a NFW halo, 
and a constant intrinsic velocity dispersion (with free parameters \lMbar, \fDM, \redisk, and \sigmaint), 
we tend to find an anti-correlation between the total baryonic mass and the dark matter fractions. 
This degeneracy is not unexpected, as it roughly traces lines of constant \vcirc, reflecting a trade-off between 
dark and baryonic matter that is accentuated by the smoothing of the different velocity curve shapes by 
the effects of beam smearing (i.e., the total mass is better constrained than the mass partitioning). 
However, the exact position and shape of this degeneracy depends on the values of other model parameters, 
particularly the intrinsic velocity dispersion \sigmaint and the disk effective radius, \redisk.

The values of \sigmaint and \redisk are connected to the dark matter fraction--baryonic mass degeneracy 
through the asymmetric drift correction (see Sec.~\ref{sec:AD_highsig}) 
and both the definition $f_{\mathrm{DM}}(\re\equiv\redisk)$ and the role of \redisk in setting the model rotation curve profile, respectively. 
In general, increasing both \sigmaint and \redisk results in shifting the region of highest \fDM--\lMbar likelihood towards 
lines of higher \vcirc, as illustrated in the left panel of Figure~\ref{fig:fdm_mbar_degen_illustration}. 
Increasing \sigmaint leads to larger asymmetric drift corrections, or lower \vrot for a fixed model \vcirc (i.e., fixed total mass). 
Matching the observed \vrot thus requires increasing \vcirc. 
When increasing \redisk, the intrinsic rotation curve profile is stretched to peak/flatten at larger radii, 
without impacting the maximum rotation velocity. 
Therefore, to match the inner rising profile of the observed $\vrot(r)$, the total mass must also be increased (i.e., higher model \vcirc). 
The exact position and extent of this degeneracy is more complex than this fixed-value illustration, 
as all four parameters (\lMbar, \fDM, \redisk, \sigmaint) are free. 
Nonetheless, this simplified picture helps to understand the likelihood degeneracy for the full fit, as a range of 
\sigmaint or \redisk values will contribute to a broadening of the degeneracy region (i.e., spanning more values of \vcirc), 
roughly corresponding to overlapping regions of fixed \sigmaint and \redisk.

The final posterior degeneracy between \fDM and \lMbar depends not only on the likelihood degeneracy, 
but also on the Gaussian priors on \lMbar and \redisk. 
In the limit where the data are not highly informative, 
the Gaussian priors will help to restrict the posteriors of these parameters to narrower ranges. 
Thus, the \lMbar prior will help break the \fDM--\lMbar degeneracy by ``picking out'' a 
subsection of the \lMbar values covered (see Figure~\ref{fig:fdm_mbar_degen_illustration}, right panel). 
The \redisk prior will also have an impact, since a narrowing of the \redisk parameter space will translate to 
narrowing of the diagonal shift of the \fDM--\lMbar anti-correlation (effectively, fewer overlapping 
degeneracy regions, as fewer \redisk values are highly probable).

To illustrate the parameter degeneracies and the role of the priors, 
in Figure~\ref{fig:PLP_corner} we show the prior, likelihood, and posterior distributions 
in all 1D and 2D projected spaces for one galaxy in our sample (BX482). 
The posterior distribution (black lines and black/grey filled regions) is determined from the 1D MCMC sampler chain, 
and the likelihood distribution (orange lines and contours) is approximated by performing a separate MCMC sampling with uninformative 
(i.e., bounded flat) priors for all parameters.\footnote{Technically, this gives a posterior distribution 
that is directly proportional to the likelihood. However, as we are concerned with 
relative distributions for this illustration, this is an acceptable approximation.  
See \citet{Hogg18} for an in-depth discussion on MCMC sampling and prior, likelihood, and posterior distributions.} 
The 1D priors are shown with dotted blue lines, and the center and standard deviation of the Gaussian priors on 
\lMbar and \redisk are marked with blue lines and shaded regions in the projected 1D and 2D panels, 
and the 1D histograms are all normalized, as the correct scaling factors have not been determined.

The impact of the Gaussian priors on \lMbar and \redisk can be seen in both the 1D and 2D histograms, 
as the posterior peaks for these parameters lie between the prior and likelihood peaks 
(as do the MAP values, which are found by jointly analyzing all free parameters; white plus signs), 
and the marginalized posterior distribution peaks are less broad than those of the likelihood. 
Because of the \fDM--\lMbar degeneracy (bottom left panel), the \lMbar prior also impacts the posterior distribution of \fDM, 
and yields a narrower posterior degeneracy than for the likelihood distribution. 
The restriction of the \redisk values from the prior also contributes to narrowing of the 
degeneracy region, but in the direction perpendicular to the anti-correlation (i.e., spanning a narrower range of constant \vcirc; see Figure~\ref{fig:fdm_mbar_degen_illustration}). 
Although the actual posterior probability distributions for each galaxy will also 
depend on other model parameters and on the peculiarities of the observed rotation and dispersion curves, 
the trends discussed here account for the overall qualitative properties 
of the posteriors determined from our MCMC analysis of the RC41 sample.

% ++++++++++++++++++++++++++++++++++++++++++++++++++++++++++++++
% ++++++++++++++++++++++++++++++++++++++++++++++++++++++++++++++

\end{document}